\documentclass[apj]{emulateapj}

\usepackage[normalem]{ulem}
\usepackage{xfrac}
\usepackage{float}
\usepackage{epsfig}
\usepackage{longtable}
\usepackage{comment}
\usepackage{amsmath}
\usepackage{mathtools}
\usepackage{threeparttablex}
\usepackage[usenames,dvipsnames]{xcolor}
\usepackage[colorlinks = true,
            linkcolor = black,
            urlcolor  = blue,
            citecolor = black,
            anchorcolor = blue]{hyperref}

\long\def\symbolfootnote[#1]#2{\begingroup%
\def\thefootnote{\fnsymbol{footnote}}\footnote[#1]{#2}\endgroup}

\def\um{$\mu$m}

\def\x{$\times$}
\def\twid{$\sim$}

\def\uJy{$\mu$Jy}
\def\figtxt{\footnotesize}

\def\th{$^{\mathrm{th}}$}

\def\z{redshift}

\def\zf{ZFOURGE}

\def\mstellar{$M_*$}

\def\msol{$M_{\odot}$}

\def\logm{log($M_*$/\msol )}

\def\qui{quiescent}

\def\lir{L$_{\mathrm{IR}}$}
\def\luv{L$_{\mathrm{UV}}$}
\def\lx{$\mathrm{L_{X}}$}
\def\lsol{L$_{\odot}$}

\def\sfrmass{SFR$-M_*$}
\def\rf{rest-frame}
\def\psiuvir{$\Psi_{\mathrm{UV+IR}}$}

\def\psiir{$\Psi_{\mathrm{IR}}$}

\begin{document}

\title{The \sfrmass\ Relation and Empirical Star-Formation Histories from ZFOURGE\footnotemark[$\ast$] at $0.5 < z < 4$}
\footnotetext[$\ast$]{This paper includes data gathered with the 6.5 meter Magellan Telescopes located at Las Campanas Observatory, Chile.}

\author{Adam R. Tomczak$^{1,2,3}$}
\author{Ryan F. Quadri$ ^{1,2}$}
\author{Kim-Vy H. Tran$^{1,2}$}
\author{Ivo Labb\'e$^4$}
\author{Caroline M. S. Straatman$^4$}
\author{Casey Papovich$^{1,2}$}
\author{Karl Glazebrook$^5$}
\author{Rebecca Allen$^{5,6}$}
\author{Gabreil B. Brammer$^7$}
\author{Michael Cowley$^{6,8}$}
\author{Mark Dickinson$^9$}
\author{David Elbaz$^{10}$}
\author{Hanae Inami$^9$}
\author{Glenn G. Kacprzak$^5$}
\author{Glenn E. Morrison$^{11,12}$}
\author{Themiya Nanayakkara$^5$}
\author{S. Eric Persson$^{13}$}
\author{Glen A. Rees$^{8}$}
\author{Brett Salmon$^{1,2}$}
\author{Corentin Schreiber$^{10}$}
\author{Lee R. Spitler$^{6,8}$}
\author{Katherine E. Whitaker$^{14,15}$\footnotetext[15]{Hubble Fellow}}

\affil{$^1$ George P. and Cynthia Woods Mitchell Institute for Fundamental Physics and Astronomy,\\Texas A\&M University, College Station, TX, 77843-4242 USA}
\affil{$^2$ Department of Physics and Astronomy, Texas A\&M University, College Station, TX, 77843-4242 USA}
\affil{$^3$ Department of Physics, University of California-Davis, One Shields Avenue, Davis, CA 95616, USA; \href{mailto:artomczak@ucdavis.edu}{artomczak@ucdavis.edu}}
\affil{$^4$ Leiden Observatory, Leiden University, P.O. Box 9513, 2300 RA Leiden, The Netherlands}
\affil{$^5$ Centre for Astrophysics \& Supercomputing, Swinburne University, Hawthorn, VIC 3122, Australia}
\affil{$^6$ Australian Astronomical Observatory, 105 Delhi Rd, Sydney, NSW 2113, Australia}
\affil{$^7$ Space Telescope Science Institute, 3700 San Martin Drive, Baltimore, MD 21218, USA}
\affil{$^8$ Department of Physics \& Astronomy, Macquarie University, Sydney, NSW 2109, Australia}
\affil{$^9$ National Optical Astronomy Observatory, 950 North Cherry Avenue, Tucson, AZ 85719, USA}
\affil{$^{10}$ Laboratoire AIM-Paris-Saclay, CEA/DSM/Irfu - CNRS - Universit\'e Paris Diderot,\\ CEA-Saclay, pt courrier 131, 91191 Gif-sur-Yvette, France}
\affil{$^{11}$ Institute for Astronomy, University of Hawaii, Manoa, Hawaii 96822-1897 USA}
\affil{$^{12}$ Canada-France-Hawaii Telescope Corp., Kamuela, Hawaii 96743-8432, USA}
\affil{$^{13}$ Carnegie Observatories, Pasadena, CA 91101, USA}
\affil{$^{14}$ Department of Astronomy, University of Massachusetts, Amherst, MA 01003, USA}

\begin{abstract}

We explore star-formation histories (SFHs) of galaxies based on the evolution of the star-formation rate stellar mass relation (\sfrmass).
Using data from the FourStar Galaxy Evolution Survey (ZFOURGE) in combination with far-IR imaging from the {\it Spitzer} and {\it Herschel} observatories we measure the \sfrmass\ relation at $0.5 < z < 4$.
Similar to recent works we find that the average infrared SEDs of galaxies are roughly consistent with a single infrared template across a broad range of redshifts and stellar masses, with evidence for only weak deviations.
We find that the \sfrmass\ relation is not consistent with a single power-law of the form $\mathrm{SFR} \propto M_*^{\alpha}$ at any redshift; it has a power-law slope of $\alpha \sim 1$ at low masses, and becomes shallower above a turnover mass (M$_0$) that ranges from $10^{9.5}$$-$$10^{10.8}$ \msol, with evidence that M$_0$ increases with \z.
We compare our measurements to results from state-of-the-art cosmological simulations, and find general agreement in the slope of the \sfrmass\ relation albeit with systematic offsets.
We use the evolving \sfrmass\ sequence to generate SFHs, finding that typical SFRs of individual galaxies rise at early times and decline after reaching a peak.
This peak occurs earlier for more massive galaxies.
We integrate these SFHs to generate mass-growth histories and compare to the implied mass-growth from the evolution of the stellar mass function.
We find that these two estimates are in broad qualitative agreement, but that there is room for improvement at a more detailed level.
At early times the SFHs suggest mass-growth rates that are as much as 10\x\ higher than inferred from the stellar mass function.
However, at later times the SFHs under-predict the inferred evolution, as is expected in the case of additional growth due to mergers.

\end{abstract}

\section{Introduction}

Over the past two decades our understanding of the buildup of stellar matter in the universe has advanced markedly through a wealth of multiwavelength galaxy surveys \citep[for a review see][]{Madau14}.
However, inferring star formation and mass growth histories of individual galaxies is a non-trivial undertaking, and a variety of methods have been used in the literature.
One class of methods involves ``archeological'' studies of nearby galaxies, either by studying resolved stellar populations or by detailed modeling of high signal-to-noise spectra \citep[e.g.][]{Dolphin03, Heavens04, Thomas05}.
However degeneracies in age, metallicity, and extinction complicate modeling with these techniques.
Furthermore, these techniques become difficult or impossible to apply at appreciable \z s.

This has provided motivation for lookback studies that utilize observed relations of galaxies at discrete epochs in the universe to infer how individual galaxies evolve.
One such type of study is to trace the mass-growth of galaxies selected in bins of constant cumulative co-moving number density \citep[e.g.][]{vanDokkum10, Papovich11, Patel13}.
This method assumes that the rank-ordering of a population of galaxies by stellar mass does not change as they evolve with time.
In reality this rank-ordering \emph{will} change due to mergers and stochastic variations in star-formation rates, but it is possible to approximately correct for these effects using an evolving number density criterion \citep{Leja13, Behroozi13}.

Another type of lookback study involves using the observed correlation between stellar mass and star-formation rate, hereafter referred to as the \sfrmass\ relation \citep[e.g.][]{Brinchmann04, Noeske07, Gilbank11, Whitaker12, Speagle14}.
By tracing along this evolving star-formation sequence it is possible to predict how galaxies should evolve due to star-formation \citep[e.g.][]{Leitner12, Speagle14}.
In general some disagreement between this approach and the number density selection (NDS) is expected since the former does not include growth due to mergers; indeed, \citet{Drory08} use this difference to derive the merger rate.
Disagreements may also be caused by systematic errors in mass and/or SFR estimates, as emphasized by \citet{Weinmann12} and \citet{Leja15}.

The most commonly used parameterization for the \sfrmass\ relation in the literature has been a power-law of the form log($\Psi$) = $\alpha$ log(\mstellar) + $\beta$ with $\alpha$ and $\beta$ representing the slope and normalization respectively.
At low stellar masses ($\lesssim 10^{10}$ \msol) this slope needs to be close to unity in order to maintain the roughly constant low-mass slope in the observed galaxy stellar mass function (SMF).
Many early studies, however, typically find a significantly shallower slope (see Table 4 of \citet{Speagle14}).
Furthermore, \citet{Leja15} argue that the sequence must also flatten at higher masses in order to be consistent with the SMF. 
Fortunately, recent new measurements of the \sfrmass\ relation find it to be more consistent with this picture \citep{Whitaker14, Lee15, Schreiber15, Tasca15}.

Many early works relied on estimating SFRs from \rf\ UV with assumed correction factors to account for extinction from dust.
The launch of the {\it Spitzer} Space Telescope \citep{Werner04} allowed us to directly probe the attenuated UV light of star-forming regions in galaxies emitted in the far-IR for statistically large samples of galaxies at $z > 1$. 
However, due to technical challenges, data quality in the far-IR were much poorer than in the optical/near-IR.
The launch of the {\it Herschel} Space Observatory \citep{Pilbratt10} expanded observational studies in the far-IR with improved data quality at longer wavelengths.
Combinations of {\it Spitzer} and {\it Herschel} data make it possible to constrain IR SEDs for large enough samples of galaxies to complement modern optical/near-IR galaxy surveys \citep[e.g.][]{Elbaz11, Wuyts11}.

We use the FourStar Galaxy Evolution Survey (\zf) in concert with deep far-IR imaging from {\it Spitzer} and {\it Herschel} to make new measurements of the \sfrmass\ relation and use this to perform an analysis of the two types of lookback studies previously mentioned.
The longer wavelength data from {\it Spitzer} and {\it Herschel} allow for robust SFR measurements 
\citep[e.g.][]{Kennicutt98, Chary01, Papovich07, Elbaz11}.
Combining this with accurate photometric \z s and deep stellar mass functions provided by \zf\ leads to improved constraints on the evolution of the \sfrmass\ relation and galaxy growth histories.
Throughout this paper we use a \citet{Chabrier03} IMF and $\Lambda$CDM cosmological parameters of $\Omega_{M} = 0.3$, $\Omega_{\Lambda} = 0.7$ and $h = 0.7$.
The symbol $\Psi$ will be used in reference to star-formation rates with subscripts to indicate how they were calculated.

\begin{figure*}
\epsfig{ file=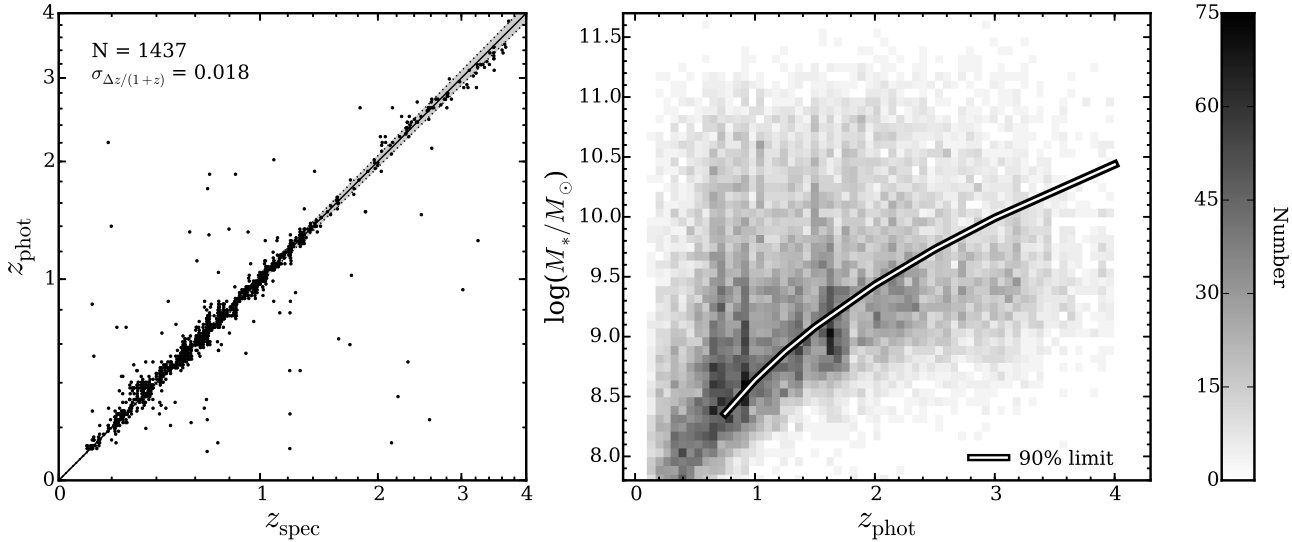 , width=0.95\linewidth }
\caption{\figtxt 
{\bf Left:} Comparison of spectroscopic to photometric \z s for 1437 objects with secure spectroscopic detections.
We estimate the NMAD scatter of $\Delta z / (1 + z_{\mathrm{spec}})$ to be 0.018 as shown by the gray shaded region with $2.7\%$ of objects being catastrophic outliers ($|\Delta z / ( 1 + z_{spec} ) | > 0.15$).
{\bf Right:} Stellar mass vs. photometric \z\ for galaxies with $\mathrm{S/N} > 5$ in the K$_s$ band.
The number of galaxies per bin is indicated by the colorbar.
Our estimated 90\% mass-completeness limit, shown by the solid line, was evaluated by estimating the distribution of $M/L$ ratios of galaxies that are slightly above the \zf\ magnitude limit K$_s=25$ and assuming the distribution is similar at the magnitude limit.
}
\label{fig:z-mass}
\end{figure*}

\section{Data and Methods}
\subsection{ZFOURGE}
\label{sec:zfourge}
\footnotetext[1]{\href{http://zfourge.tamu.edu}{http://zfourge.tamu.edu}}

The FourStar Galaxy Evolution Survey (ZFOURGE\footnotemark[1]: Straatman et al. submitted) is a deep near-IR survey conducted with the FourStar imager \citep{Persson13} covering one 11\arcmin \x 11\arcmin\ pointing in each of the three legacy fields CDF-S \citep{Giacconi02}, COSMOS \citep{Capak07} and UDS \citep{Lawrence07} reaching depths of \twid 26 mag in $J_1$, $J_2$, $J_3$, and \twid 25 mag in $H_s$, $H_l$, and $K_s$ (5$\sigma$ in $d$$=$$0\overset{''}{.}6$ apertures).
The medium-bandwidth filters utilized by this survey offer spectral resolutions $\lambda / \Delta \lambda \approx 10$, roughly twice that of their broadband counterparts.
This increase provides for finer sampling of the Balmer/4000\AA\ spectral break at $1 < z < 4$, leading to well-constrained photometric \z s.
In combination with ancillary imaging, the full photometric dataset covers the observed 0.3$-$8\um\ wavelength range.

\subsection{Redshifts and Stellar Masses}
\label{sec:eazyfast}

Photometric redshifts and rest-frame colors were measured using the public SED-fitting code EAZY \citep{Brammer08} on PSF-matched optical-NIR photometry.
EAZY utilizes a default set of six spectral templates that include prescriptions for emission lines derived from the PEGASE models \citep{Fioc97} plus an additional dust-reddened template derived from the \citet{Maraston05} models.
Linear combinations of these templates are fit to the $0.3-8$\um\ photometry for each galaxy to estimate redshifts.

A comparison of our derived photometric \z s to a sample of 1437 galaxies with secure spectroscopic \z s is shown in Figure \ref{fig:z-mass}.
We calculate a scatter of $\Delta z \: / \: (1 + z_{\mathrm{spec}}) = 1.8\%$ at $z < 1.5$ and fraction of catastrophic outliers ($|\Delta z / ( 1 + z_{spec} ) | > 0.15$) of $2.7\%$.
At $z > 1.5$ these rise to $2.2\%$ and $9\%$ respectively.
An additional analysis of $z_{\mathrm{phot}}$ accuracy can be found in Section 2 of \citet{Kawinwanichakij14} and Straatman et al. (submitted).
Spectroscopic \z s from CDF-S are taken from \citet{Vanzella08}, \citet{LeFevre05}, \citet{Szokoly04}, \citet{Doherty05}, \citet{Popesso09}, and \citet{Balestra10}.
For COSMOS spectroscopic \z s come from \citet{Lilly09} and \citet{Trump09}.
Spectroscopic \z s for UDS come from \citet{Simpson12} and \citet{Smail08}.

Stellar masses were derived by fitting stellar population synthesis templates to the $0.3-8$\um\ photometry using the SED-fitting code FAST \citep{Kriek09}.
FAST was run using a grid of \citet{Bruzual03} models assuming a \citet{Chabrier03} IMF and solar metallicity.
Exponentially declining star-formation histories ($\Psi \propto e^{-t / \tau}$) are used with log($\tau$/yr) ranging between $ 7-11$ in steps of 0.2 and allowing log(age/yr) to vary between $7.5-10.1$ in steps of 0.1.
A \citet{Calzetti00} extinction law is also incorporated with values of A$_V$ varying between $0-4$ in steps of 0.1.

Mass-completeness limits are estimated using a method similar to \citet{Quadri12}.
Briefly, we estimate the distribution of mass-to-light ratios of galaxies that are somewhat above our K$_s=25$ magnitude limit, and use this distribution to estimate the 90\% mass-completeness limit of galaxies at K$_s=25$.
These mass-completeness limits are shown in Figure \ref{fig:z-mass} along with the distribution of stellar masses and \z s of galaxies in the \zf\ catalogs.
A more complete discussion of the mass-completeness limits will be presented by Straatman et al. (submitted).

\subsection{Far-Infrared Imaging}
\label{sec:firphot}

We make use of {\it Spizer}/MIPS (GOODS-S: PI Dickinson, COSMOS: PI Scoville, UDS: PI Dunlop) and {\it Herschel}/PACS data \citep[GOODS-S:][COSMOS \& UDS: PI Dickinson]{Elbaz11} for measuring total infrared luminosities (\lir) to derive SFRs.
Imaging from these observatories used in this study include 24, 100 and 160\um.
Median 1$\sigma$ flux uncertainties for CDF-S/COSMOS/UDS are approximately 3.9/10.3/10.1 \uJy\ in the 24\um\ imaging, 0.20/0.43/0.45 mJy in the 100\um\ imaging and 0.35/0.70/0.93 mJy in the 160\um\ imaging respectively.

Due to the large PSFs of the MIPS/PACS imaging (FWHM $\gtrsim$ 4\arcsec) source blending is a considerable effect.
Therefore we use the Multi-Resolution Object PHotometry oN Galaxy Observations (MOPHONGO) code written by I. Labb\'e to extract deblended photometry in these far-IR data \citep[for a detailed discussion see][]{Labbe06, Wuyts07}.
The algorithm uses higher resolution imaging to generate a segmentation map containing information on the locations, sizes and extents of objects.
In this work we use deep K$_s$ band as the prior (FWHM $=$ 0.46\arcsec).
Point-sources coincident in both images are used to construct a convolution kernel that maps between the high and low resolution PSFs.
Objects used to construct this kernel need to be hand selected as many point-sources in the $K_s$ imaging are frequently undetected at far-IR wavelengths.
A model of each far-IR image is generated by convolving the high-resolution segmentation map with the corresponding kernel allowing the intensities of individual objects to vary freely.
Background and RMS maps are generated locally for each object on scales that are three times the 30\arcsec\ tile-size used.
By subtracting the modeled light of neighboring sources, ``cleaned'' image tiles of individual objects are produced which will be used in the stacking analysis discussed in the following section.

\begin{figure*}
\epsfig{ file=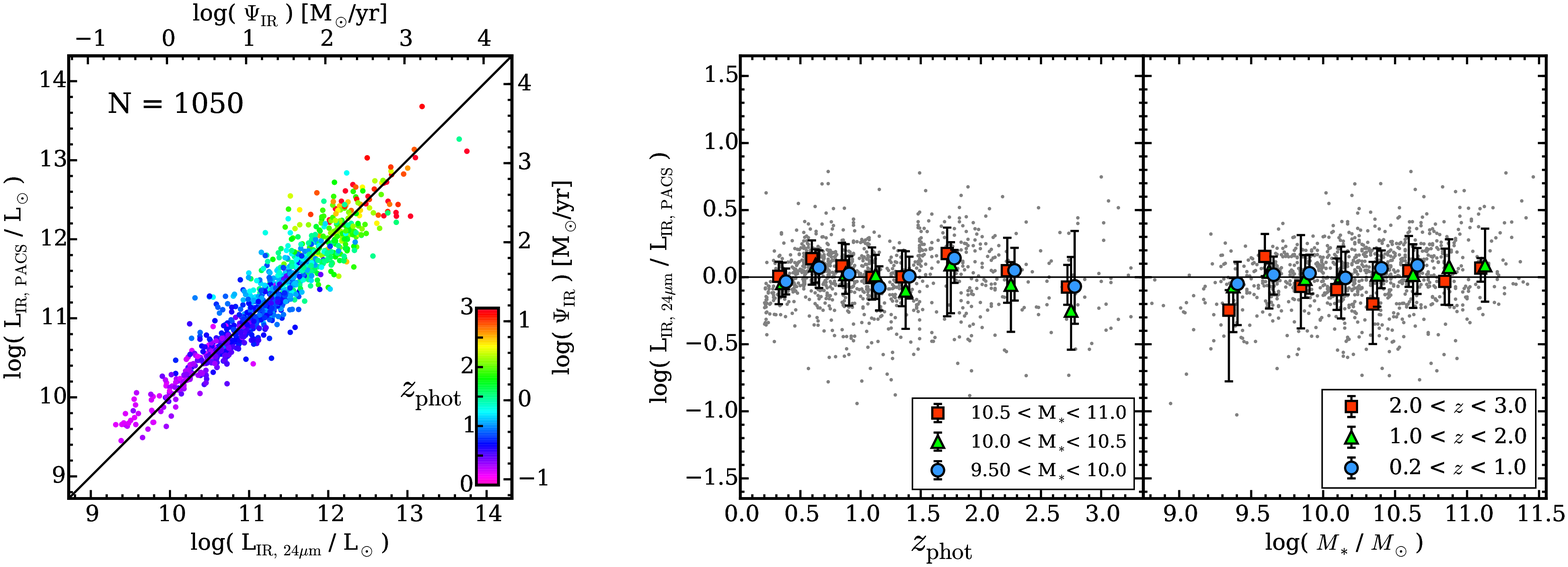 , width=0.95\linewidth }
\caption{\figtxt 
Comparison between the estimated \rf\ \lir($8-1000$\um) based on MIPS-only versus PACS-only data.
\lir\ is obtained by scaling the IR template presented by \citet{Wuyts08} to the MIPS 24\um\ data (L$_{\mathrm{IR, \, 24 \mu m}}$) or PACS 100 and 160\um\ data (L$_{\mathrm{IR, \, PACS}}$) respectively for individually detected galaxies (see Section \ref{sec:sfrmass} for more details).
Only galaxies with $\mathrm{S/N} > 3$ in all three bandpasses are considered for this comparison.
In general we obtain consistent estimates for \lir\ across a broad range of \z\ and stellar mass as shown in the two {\bf right} panels (m$_{\star} \equiv \mathrm{log}( M_* / M_{\odot} )$).
}
\label{fig:lir-compare}
\end{figure*}

\begin{figure*}[t]
\epsfig{ file=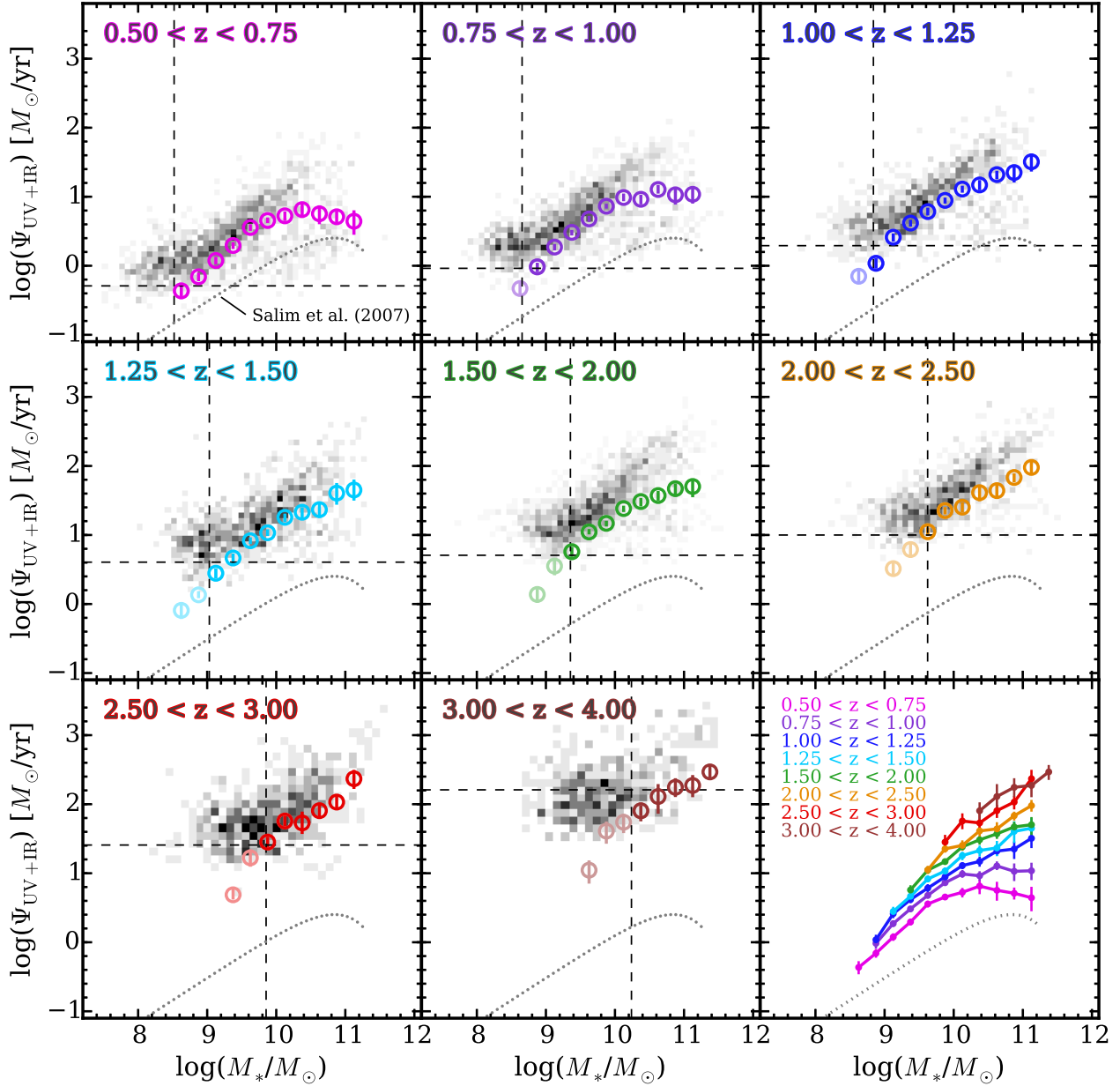 , width=0.95\linewidth }
\caption{\figtxt 
Star-formation rate vs. stellar mass relations for all galaxies.
Colored points represent stacked SFRs whereas the gray-scale shows the distribution of 24\um-detected galaxies ($\equiv$ S/N$_{24\mu\mathrm{m}} > 1$).
Vertical and horizontal dashed lines show estimated mass-completeness limits and median 1$\sigma$ 24$\mu$m flux uncertainties respectively.
The final panel shows all stacked measurements above the estimated mass-completeness limits.
The dotted line is the $z \approx 0.1$ measurement from \citet{Salim07}.
}
\label{fig:sfr-mass}
\end{figure*} 

\begin{figure*}[t]
\epsfig{ file=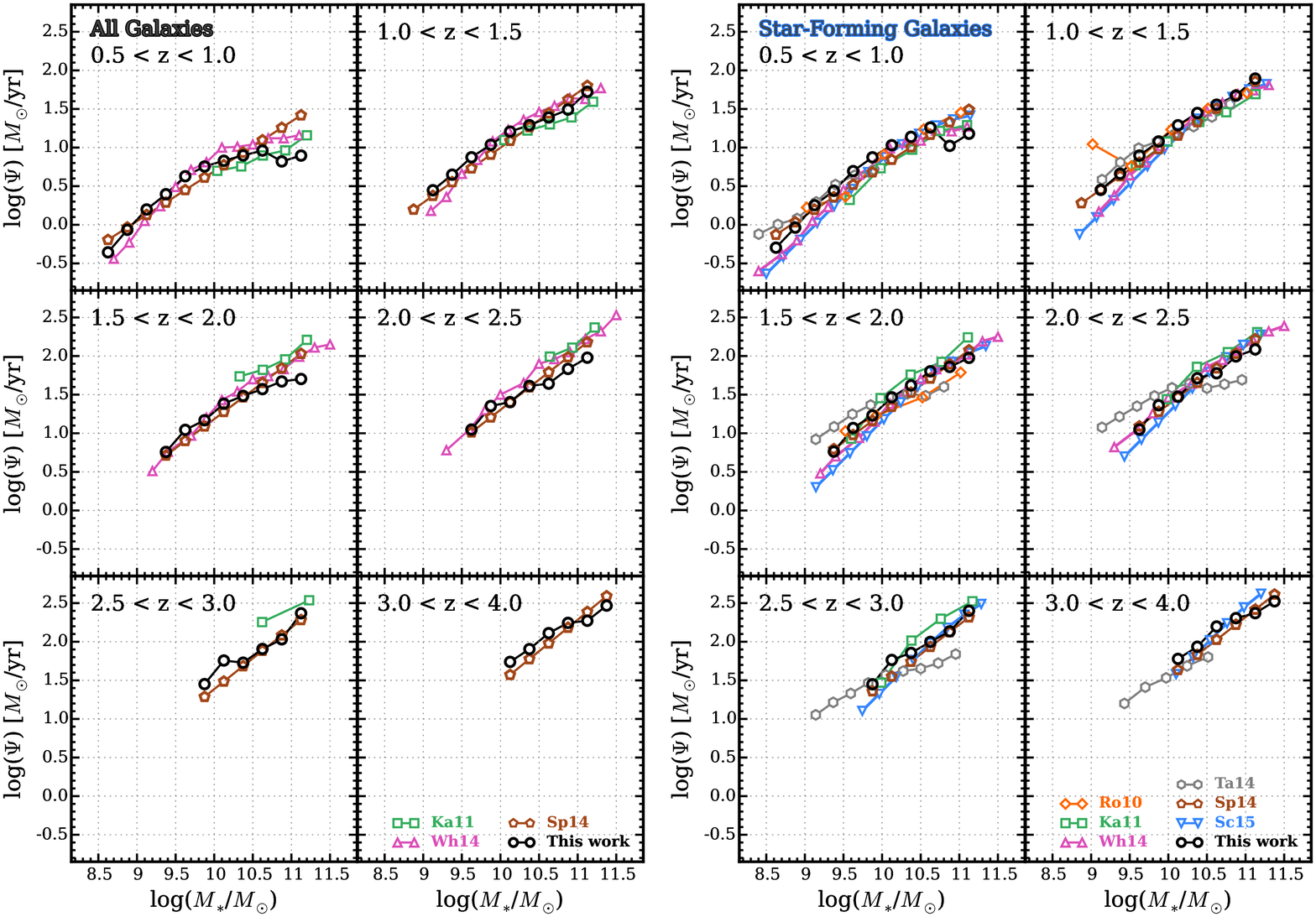 , width=0.95\linewidth }
\caption{\figtxt 
Comparison of our \sfrmass\ relations to recent measurements from literature: \citet[Ro10]{Rodighiero10}, \citet[Ka11]{Karim11}, \citet[Wh14]{Whitaker14}, \citet[Ta14]{Tasca15}, \citet[Sp14]{Speagle14}, and \citet[S15]{Schreiber15}.
Panels on the {\bf left} correspond to the full galaxy sample whereas panels on the {\bf right} correspond to actively star-forming galaxies.
All sequences shown here have been converted to a \citet{Chabrier03} IMF.
Note, due to the large \z\ bins used by \citet{Tasca15} we show weighted averages of their measurements here.
Curves for the \citet{Schreiber15} relations come from their parameterization evaluated at the central \z\ of each bin shown here (see their Section 4.1).
Similarly, the curves for \citet{Speagle14} correspond to parameterizations from their table 9: ``All'' and ``Mixed'' for the all- and star-forming galaxy samples shown here.
We also note that the \z\ bins of the \citet{Karim11} relations are different than those indicated at the top: $0.6 < z < 0.8$, $1.0 < z < 1.2$, $1.6 < z < 2.0$, $2.0 < z < 2.5$, and $2.5 < z < 3.0$ respectively.
}
\label{fig:sfr-mass_literature}
\end{figure*}

\begin{figure*}[t]
\epsfig{ file=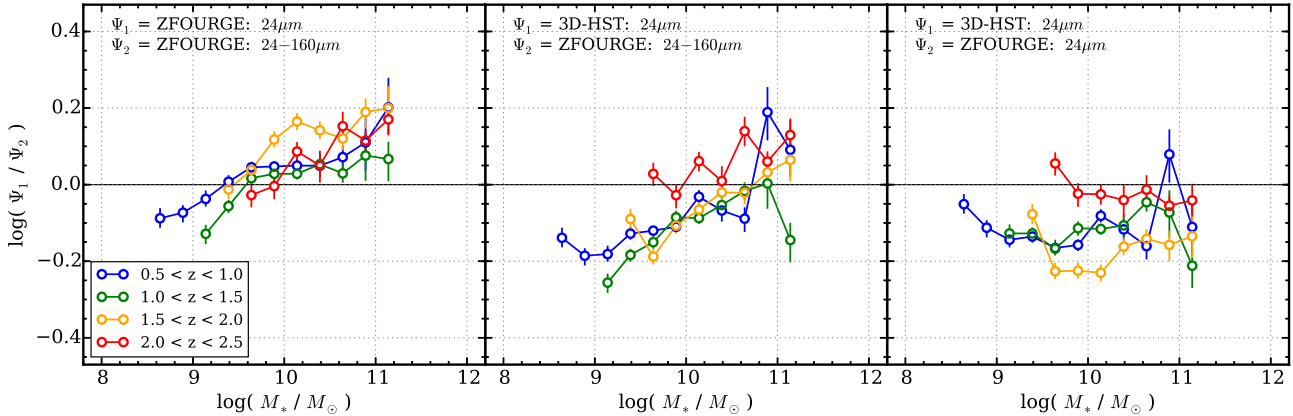 , width=0.95\linewidth }
\caption{\figtxt 
Comparison of stacked UV+IR star-formation rates using 24$-$160\um\ vs. 24\um$-$only photometry. 
Only star-forming galaxies are considered for this comparison.
The {\bf left} panel shows an internal comparison from the present dataset and reveals a clear trend where the estimated star-formation rates of more massive galaxies (typically with higher \lir) decrease when including PACs photometry.
This result is at odds with our findings for individually FIR-detected galaxies (see Fig. \ref{fig:lir-compare}) implying that galaxies with low \lir\ have different infrared SEDs than galaxies with high \lir.
The {\bf middle} panel shows a similar systematic trend for an external comparison of our UV+IR SFRs (24$-$160\um) to those of 3D-HST \citep{Whitaker14} which only utilize 24\um\ photometry in the IR.
Finally, the panel on the {\bf right} shows the comparison of 24\um-only UV+IR star-formation rates.
Given that the 3D-HST and \zf\ surveys share many similarities (see Section \ref{sec:literature} for details) these differences are indicative of the minimal systematic differences that can be expected in inter-survey comparisons.
}
\label{fig:sfr-mass_literature_residuals}
\end{figure*}

\begin{figure*}[t]
\epsfig{ file=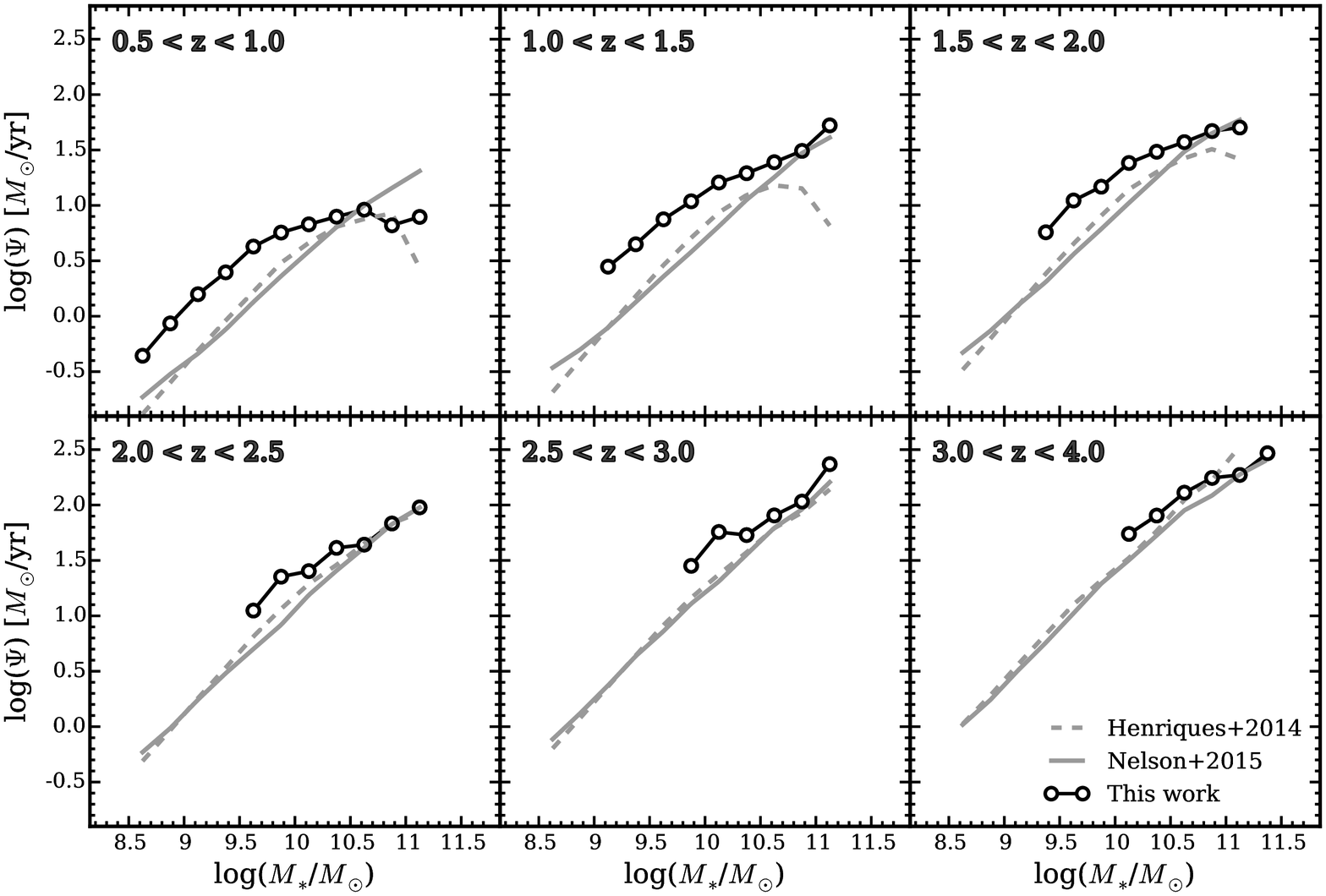 , width=0.95\linewidth }
\caption{\figtxt 
Comparison of our measured \sfrmass\ relations of all galaxies (black points) to those of recent cosmological simulations.
The solid and dashed gray lines show mean star-formation rates in bins of stellar mass from the Illustris simulation \citep{Nelson15arxiv} and the Munich galaxy formation model \citep{Henriques14arxiv} respectively.
At \logm\ $< 10.5$ the simulations produce \sfrmass\ relations with a very similar slope to the observations, although with a distinct offset to lower SFR at fixed stellar mass.
Similar to \citet{Sparre15}, we find this offset ranges from 0.17 to 0.45 dex and decreases with \z.
}
\label{fig:sfr-mass_literature_simulations}
\end{figure*}

\begin{figure*}[t]
\epsfig{ file=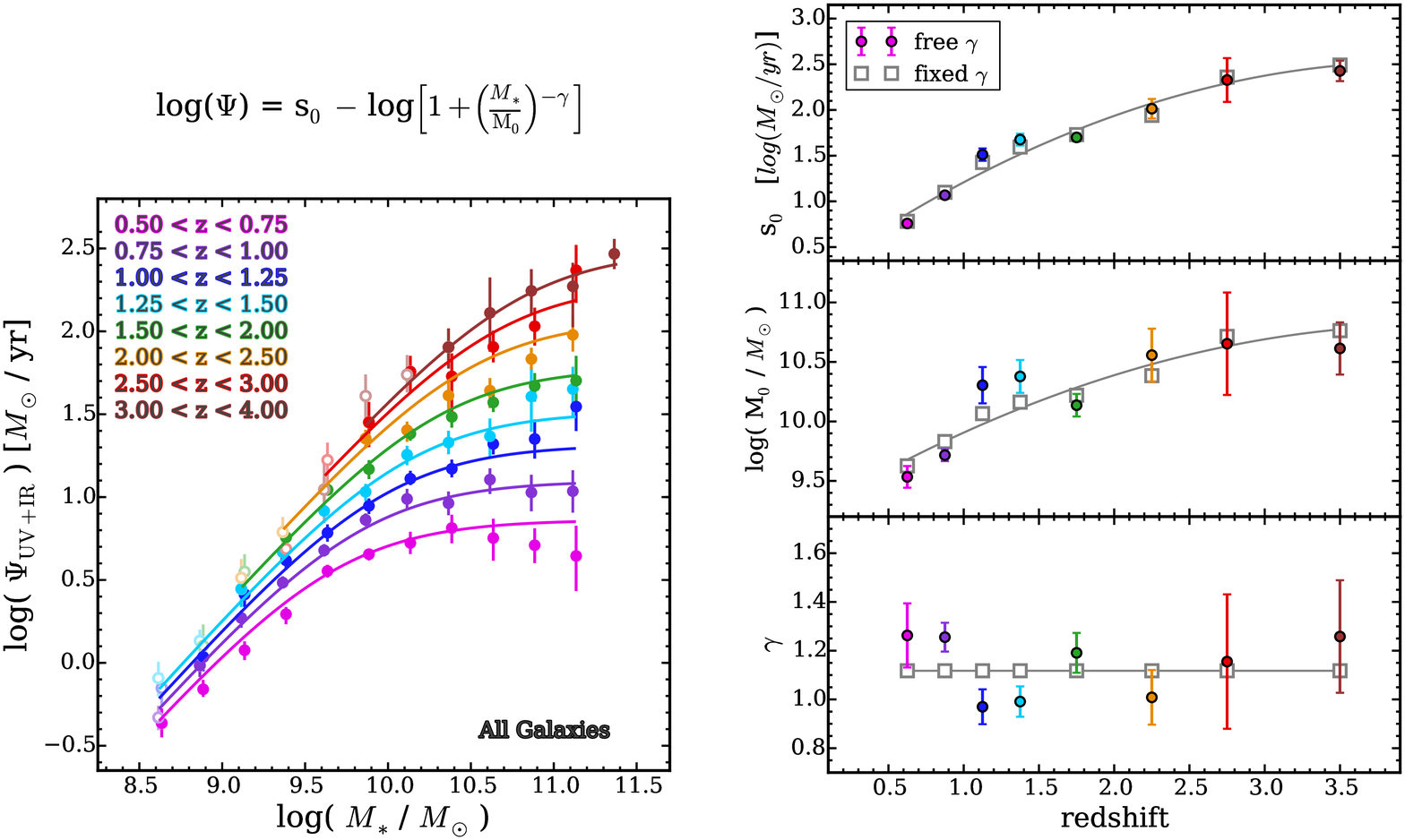 , width=0.95\linewidth }
\caption{\figtxt
Our parameterization for the \z\ evolution of the \sfrmass\ sequence of all galaxies.
We start with Equation \ref{eq:sfrmass} (shown in the upper left) which is described by three free parameters: a power-law of slope $\gamma$ at low masses which asymptotically approaches a peak star-formation rate s$_0$ at high masses with M$_0$ being the transition mass between the two behaviors.
On the {\bf right} we show these best-fitting parameters vs. \z\ and results from the fitting procedure described in Section \ref{sec:sfrmass}.
The panel on the {\bf left} shows the the corresponding redshift-parameterized \sfrmass\ relations at the central \z\ of each bin with points showing the stacked measurements from Figure \ref{fig:sfr-mass}. 
}
\label{fig:sfrmass-parameterization_TOT}
\end{figure*}

\begin{figure*}[t]
\epsfig{ file=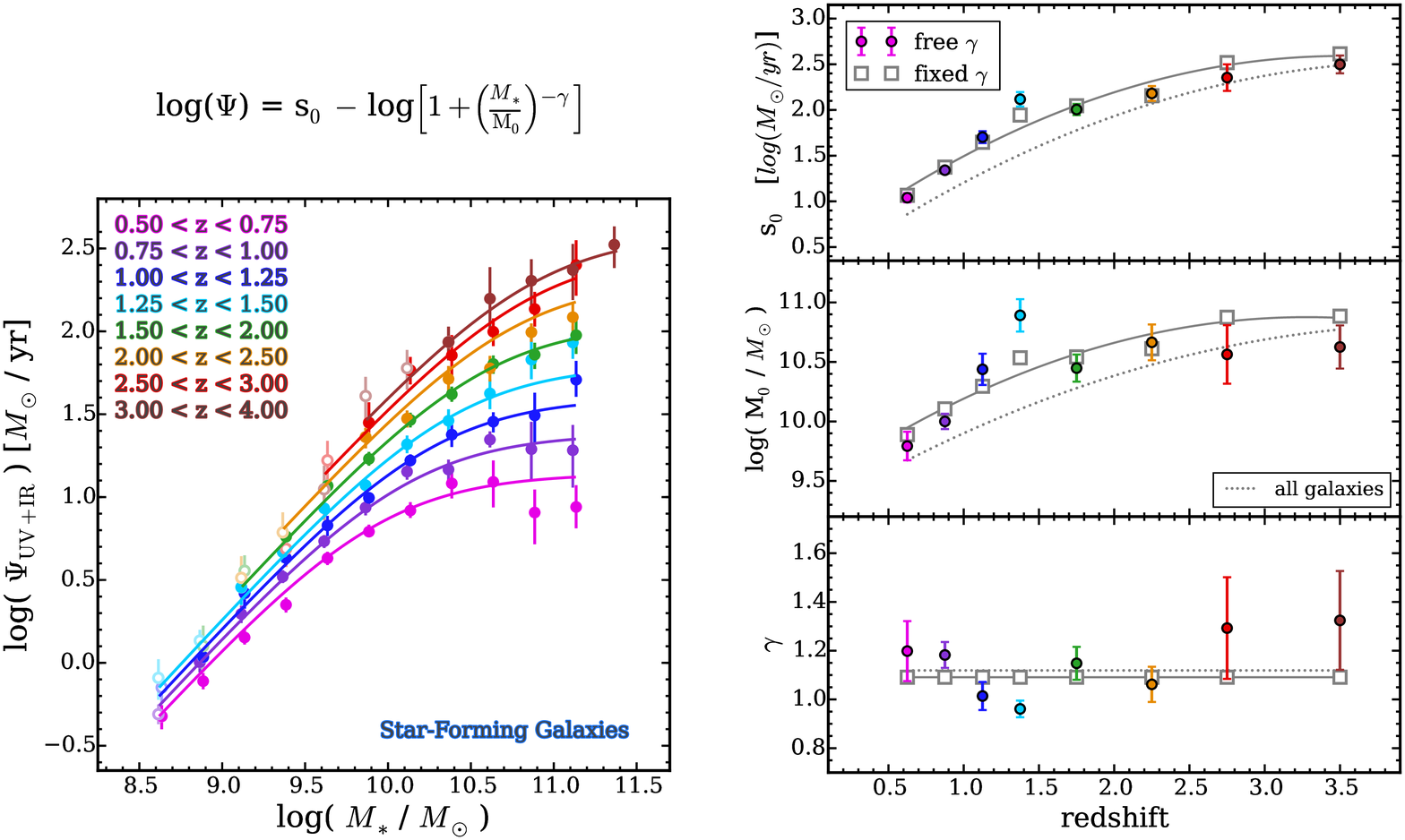 , width=0.95\linewidth }
\caption{\figtxt
Same as Figure \ref{fig:sfrmass-parameterization_TOT} but for UVJ-selected star-forming galaxies only.
Evolution in M$_0$ is still apparent indicating that this behavior is not driven exclusively by the buildup of massive \qui\ galaxies at low \z.
Differences between these \sfrmass\ relations and those for all galaxies in the previous figure are roughly consistent with the evolution of the stellar mass functions of star-forming and quiescent galaxies from the ZFOURGE survey as measured by \citet{Tomczak14}.
}
\label{fig:sfrmass-parameterization_SF}
\end{figure*}

\begin{figure}[t]
\epsfig{ file=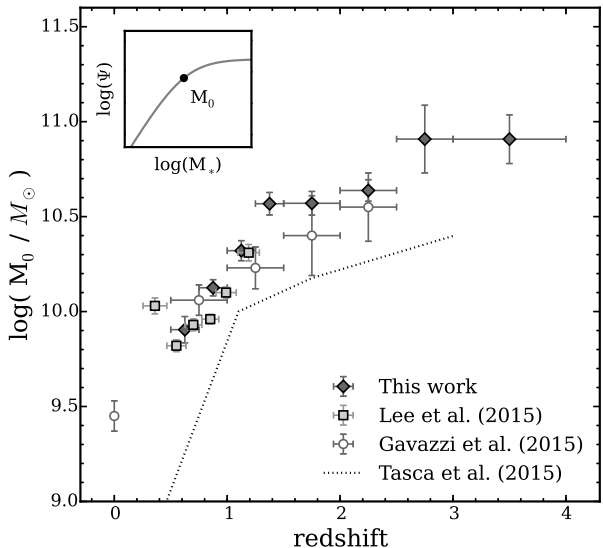 , width=0.95\linewidth }
\caption{\figtxt
Evolution of the turnover mass (M$_0$) in the \sfrmass\ relation of star-forming galaxies.
An illustration of the location of M$_0$ as defined in equation \ref{eq:sfrmass} is shown in the upper left.
Results shown from this work are from the ``fixed $\gamma$" fitting procedure.
The statistical significance of our measured correlation is markedly high (Pearson correlation coefficient of $r = 0.92$).
Also shown are recent measurements from literature.
There is excellent agreement at the overlapping \z s between our measurements and those of \citet{Lee15} and \citet{Gavazzi15}.
We do observe a roughly uniform offset of $\approx 0.5$ dex with \citet{Tasca15} which may be the result of a different parameterization used to measure the turnover mass.
}
\label{fig:M0_evolution}
\end{figure}

\begin{figure*}
\epsfig{ file=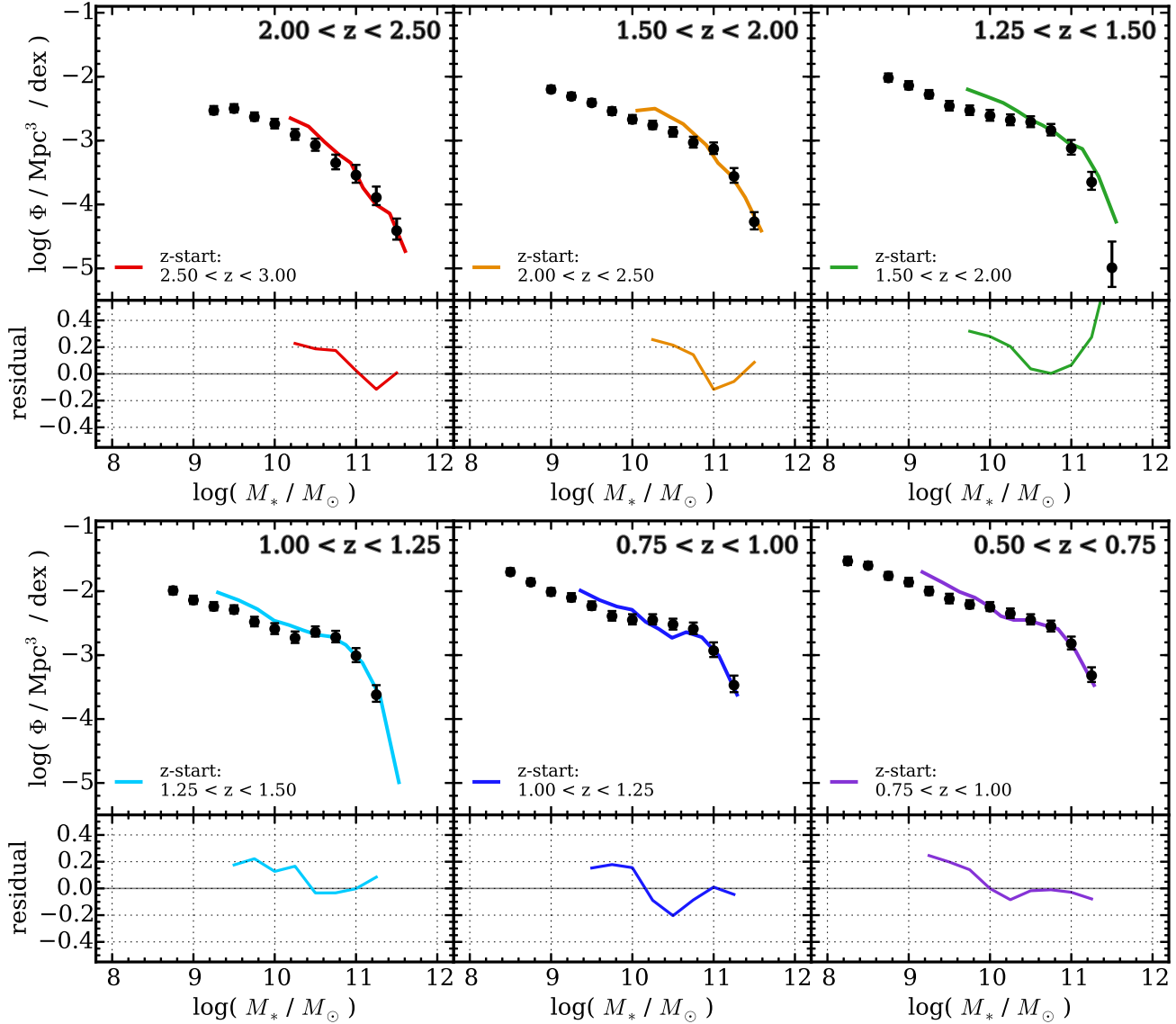 , width=0.95\linewidth }
\caption{\figtxt 
Implied growth of the galaxy stellar mass function due to star-formation.
Each panel shows the observed SMF from \citet{Tomczak14} at the \z s indicated in the upper-right corner.
Curves in each panel represent the SMF from the preceding \z\ bin evolved forward in time based on our parameterized \sfrmass\ relation for all galaxies.
Redshifts of the original SMFs (i.e. ``starting'' \z s) are indicated in the legend.
Residuals between the evolved and observed SMFs for each \z\ bin are shown in the lower panels.
We observe that the numbers of galaxies at $M_* < 10^{10.5}$ \msol\ are consistently overproduced at each \z\ by $\approx 0.2-0.3$ dex. 
It is important to note that galaxy merging is not accounted for in the inferred SMFs, thus, at least part of this offset must be caused by this effect.
However this would require between $25-65$\% of these galaxies to merge with a more massive galaxy per Gyr, which substantially exceeds current estimates of galaxy merger rates \citep[e.g.][]{Lotz11, Williams11, Leja15}.
}
\label{fig:smf-evo}
\end{figure*}

\begin{figure*}[t]
\epsfig{ file=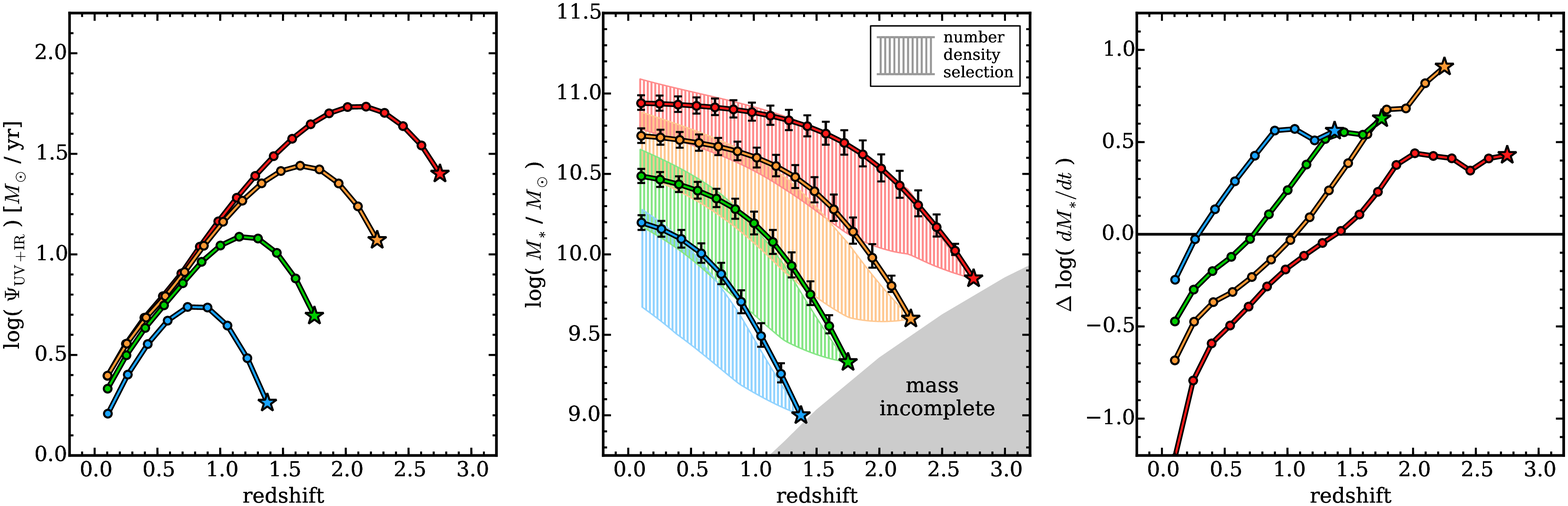 , width=0.95\linewidth }
\caption{\figtxt 
Differential SFHs ({\bf left}) and mass-growth profiles ({\bf middle}) extracted from the evolution of the \sfrmass\ relation of all galaxies.
To extend these measurements to $z \approx 0.1$ we use the \sfrmass\ relation of \citet{Salim07}.
Star symbols indicate four arbitrary sets of initial conditions consistent with Equations \ref{eq:sfrmass} and \ref{eq:parameterization_tot} which are then propagated forward in time.
Mass loss due to stellar evolution is accounted for according to Equation 16 of \citet{Moster13}.
Hatched regions in the {\bf middle} panel show the inferred growth profiles generated by mapping the predicted number density evolution \citep{Behroozi13} to the galaxy SMF as a function of \z.
Although there is broad agreement between these two techniques, we note that the integrated mass-growth curves are more accelerated, growing more rapidly at early times and slowing to lesser rates at later times.
This is shown clearly in the {\bf right} panel which plots the difference between the time derivatives of these two approaches.
We note, however, that this is only a comparison of the median evolution and ignores the $\pm 1\sigma$ scatter.
}
\label{fig:sfhs}
\end{figure*}

\subsection{Sample Selection and Stacking}
\label{sec:sample}

Modern near-infrared galaxy surveys have made it possible to detect approximately mass-complete samples of galaxies to high \z s ($z \approx 4$).
Unfortunately however, imaging used to probe obscured star-formation (typically far-IR and radio) rarely ever reach complementary depths.
Thus, many studies over the past several years have turned to measuring SFRs from stacked data in order to compensate for this disparity \citep[e.g.][]{Dunne09, Rodighiero10, Karim11, Whitaker14, Schreiber15}. 
However it is important to keep in mind that the interpretation of stacked results may be complicated by the fact that the intrinsic distribution of SFRs may not be unimodal or symmetric.

We classify galaxies as either actively star-forming or quiescent using the UVJ color-color diagram \citep{Labbe05, Wuyts07, Williams09}.
The rest-frame (U$-$V) and (V$-$J) colors are estimated using EAZY (Section \ref{sec:eazyfast}).
The advantage of this diagram is that it effectively separates the two reddening vectors caused by aging and dust extinction, decreasing the likelihood of dust-enshrouded star-forming galaxies being identified as \qui.
The UVJ diagram is thus a more effective tool for categorizing galaxies into star-forming and \qui\ subsamples than a simple color-magnitude criterion.

The deep near-IR photometry (K$_s \approx 25$) of ZFOURGE allows us to reliably select galaxies based on stellar mass.
Across the entire \z\ range considered in this work ($0. 5 < z < 4$) we detect 12,433 galaxies in the K$_s$ band imaging that lie above our estimated mass-completeness limits.
From this mass-complete sample, we find that 5,875 (47\%), 8,542 (69\%) and 8,630 (69\%) are not detected in the 24, 100 and 160\um\ images respectively (where detection is defined as S/N $>$ 1).
As such, we resort to stacking of the far-IR photometry for our K$_s$-selected sample in order to more precisely measure fluxes for ensembles of galaxies.
In bins of \z\ and stellar mass, we average-combine ``cleaned'' image tiles (see Section \ref{sec:firphot}) of individual galaxies for each of the far-IR bandpasses.
Stacking of ``cleaned'' imaging has been shown to significantly decrease contamination from blended sources \citep[see also][]{Fumagalli14, Whitaker14}.
Finally, photometry is measured in apertures of 3.5\arcsec, 4.0\arcsec and 6.0\arcsec\ for the 24, 100 and 160\um\ respectively with a background subtraction as measured from an annulus of radii 15\arcsec$-$19\arcsec\ on each stack.
PSFs generated from bright objects are used to derive aperture corrections of 2.21, 1.76, and 1.61 respectively for the 24, 100 and 160\um\ imaging.
Because these PSFs were constructed on the same 30'' tile-size these are not corrections to total flux, thus, we adopt additional correction factors of 1.2, 1.38, and 1.54 to account for flux that falls outside the tile.

In order to estimate the error on the mean flux measured for each stack we perform 100 bootstrap resamplings on each mass-\z\ subsample.
Stacks of the 24, 100, and 160\um\ tiles are generated from these resamplings from which fluxes are measured as described above.
We take the inter-68\th\ percentile of these flux distributions as the corresponding uncertainties on the estimated IR fluxes.
In Section \ref{sec:sfrmass} we describe how these uncertainties are propagated to estimates of \lir.
It is worth mentioning that fluxes measured from stacking are subject to biases due to the clustering of galaxies. 
However, detailed simulations have shown this effect to be negligible at the image resolution of our dataset \citep{Viero13, Schreiber15} and using ``cleaned'' image tiles also helps to minimize contamination from neighboring sources.

We remove sources suspected of hosting active galactic nuclei (AGN) from all samples based on radio, X-ray and IR indicators.
Radio AGN are identified as sources with 1.4 GHz excess having $\Psi_{\mathrm{1.4}} / \Psi_{\mathrm{IR}} \geq 3$ where $\Psi_{\mathrm{1.4}}$ is the radio-inferred SFR based on equation 6 of \citet{Bell03} and \psiir\ is the IR-inferred SFR discussed in Section \ref{sec:sfrs}.  
More discussion of the selection and properties of radio AGN will be provided in Rees et al. (submitted).
Unobscured X-ray AGN are identified as having $10^{42} \leq \mathrm{L_{X}} \leq 10^{44}$ and $\mathrm{HR} < -0.2$ where \lx\ and HR are the \rf\ X-ray luminosity in erg/s and hardness ratio respectively.
All objects with $\mathrm{L_{X}} > 10^{44}$ are classified as QSOs and also rejected.
Infrared AGN are identified based on an adaptation from the criteria of \citet{Messias12} and will be presented in more detail by Cowley et al. (submitted).

\subsection{Star-Formation Rate Measurements}
\label{sec:sfrs}

We calculate total star-formation rates by adding contributions from UV and IR light.
This approach assumes that the IR emission of galaxies (\lir) originates from dust heated by the obscured UV light of young, massive stars.
Thus by adding its contribution to that of the unobscured UV luminosity (\luv) the total SFR for galaxies can be calculated. 
We use the conversion from \citet{Bell05} scaled to a \citet{Chabrier03} IMF to derive SFRs from our data:

\begin{equation}
\Psi_{\mathrm{UV+IR}} \: [M_{\odot} / \mathrm{yr}]   =   1.09 \times 10^{-10} \: ( \mathrm{L}_{\mathrm{IR}}
+  2.2 \mathrm{L}_{\mathrm{UV}} )
\end{equation} \\[-3mm]

\noindent
where \lir\ is the integrated 8$-$1000\um\ luminosity and \luv $\, =$$\, 1.5 \, \nu \, \mathrm{L}_{\nu, 2800}$ represents the \rf\  1216$-$3000\AA\ luminosity, both in units of \lsol.

For our stacking analysis we aim to measure the average star-formation rate of galaxies in bins of \z\ and stellar mass.
In each mass-redshift bin we use the median \rf\ 2800\AA\ luminosity output by EAZY for \luv\ and estimate 1$\sigma$ uncertainties from 100 bootstrap resamplings.
We estimate bolometric infrared luminosities ($\mathrm{L_{IR} \equiv L_{8-1000\mu m}}$) by fitting an IR spectral template to the stacked 24-160\um\ photometry.
The template introduced by \citet{Wuyts08}, hereafter referred to as the W08 template, was constructed by averaging the logarithm of the spectral template library from \citet{Dale02} motivated by results from \citet{Papovich07}.
The validity of this luminosity-independent conversion has been demonstrated by \citet{Muzzin10} through comparison SFRs derived from H$\alpha$ versus 24\um\ fluxes for a sample of galaxies at $z \sim 2$.

Furthermore, \citet{Wuyts11} find that at $0 < z < 3$ this luminosity-independent conversion yields consistent \lir\ estimates from 24\um\ when compared to \lir\ derived from PACS photometry from the {\it Herschel} PEP survey \citep{Lutz11}.
For our stacking analysis, we smooth the W08 template by the \z\ distribution of the galaxies in each mass-\z\ bin prior to fitting.

Errors on \lir\ were estimated from 100 Monte Carlo simulations of the stacked IR fluxes.
For each mass-redshift bin we perturb the stacked $24-160$\um\ fluxes by a normal probability density function (PDF) of width given by the estimated uncertainties described in Section \ref{sec:sample}.
Infrared luminosities are calculated for each iteration in the same way as described above.
Errors for \lir\ are derived from the 68\th\ percentile range of each \lir\ distribution.
We combine these with the uncertainties estimated for \luv\ to derive uncertainties on \psiuvir.
All measurements of UV and IR luminosities, star-formation rates, and corresponding errors can be found in Table 1.

We test the W08 template against the present dataset using a sample of 1050 well-detected galaxies ($\mathrm{S/N} > 3$ in all FIR bands).
For each galaxy we fit the W08 template separately to the MIPS (24\um) and the PACS (100+160\um) photometry.
In Figure \ref{fig:lir-compare} we show a comparison between the MIPS-only and PACS-only cases.
Although we observe a general scatter of \twid 0.2 dex we find an overall consistency with no dominant systematic trends.
Even when subsampling in \z\ and stellar mass the mean offset is nearly always within the scatter.
We do note, however, the presence of a weak systematic trend with \z\ in the middle panel of Figure \ref{fig:lir-compare} which is likely caused by PAH features shifting through the MIPS 24\um\ passband.
Nevertheless, the consistency between the MIPS-only and the PACS-only estimates of \lir\ suggests that the W08 template effectively describes the average IR SED of galaxies.
It further suggests that reasonably reliable SFR estimates can be made with just a single IR band.
In this section we have focused only on galaxies that are individually detected in the FIR bands.
In section 3.1 we present evidence that the systematic errors become slightly larger for our fainter stacked samples.

\section{The \sfrmass\ Relation}
\label{sec:sfrmass}

In Figure \ref{fig:sfr-mass} we show our measurements of the \sfrmass\ relation for all galaxies in eight \z\ bins spanning $0.5 < z < 4$.
Evaluating completeness limits for \psiuvir\ is complicated since the depth of the far-IR imaging in CDF-S is deeper than in COSMOS and UDS.
Furthermore, the ratio of IR to UV flux (infrared excess: IRX $\equiv$ \lir/\luv) is strongly correlated with mass \citep[e.g.][]{Papovich06, Whitaker14}, therefore, the completeness in \psiuvir\ will also be a function of stellar mass.
Thus, to provide a visual guide in Figure \ref{fig:sfr-mass} we plot the 1$\sigma$ MIPS 24\um\ flux uncertainty converted to \psiir\ as horizontal dashed lines.
Due to the different depths in the three fields we use the average of the estimated 1$\sigma$ flux variances in each of the fields for this conversion: $\sfrac{1}{3} \sqrt{3.9^2 + 10.3^2 + 10.1^2} = 6$\uJy.
We then scale the W08 template to this flux value, shifted to the upper \z\ of each bin to calculate approximate limiting SFRs.
Errors on stacked SFRs are determined from 100 bootstrap resamplings of their respective UV+IR stacks.
The 68th-percintile range from the resulting distribution of SFRs is used to derive the 1$\sigma$ error (see Table \ref{tab:sfrmass}).

\subsection{Comparison to Literature}
\label{sec:literature}

In Figure \ref{fig:sfr-mass_literature} we compare our \sfrmass\ relations to recent results from the literature.
The chosen \sfrmass\ relations come from \citet{Rodighiero10}, \citet{Karim11}, \citet{Whitaker14}, \citet{Tasca15}, \citet{Speagle14}, and \citet{Schreiber15}.
All works have been scaled to a \citet{Chabrier03} IMF for consistency.
Overall there is good agreement among all of the measurements presented in Figure \ref{fig:sfr-mass_literature}.  
For the full sample of galaxies (star-forming plus quiescent), the median inter-survey discrepancy at fixed stellar mass is 0.2, 0.17, 0.15, 0.16, 0.21, and 0.13 dex in the \z\ bins ($0.5 < z < 1$), ($1 < z < 1.5$), ($1.5 < z < 2$), ($2 < z < 2.5$), ($2.5 < z < 3$), and ($3 < z < 4$) respectively.
For the star-forming sample these median discrepancies are 0.24, 0.17, 0.32, 0.17, 0.30, and 0.17 dex respectively.
These differences are consistent with the inter-publication scatter found by \citet{Speagle14} which draws on a larger sample of published \sfrmass\ relations.
Furthermore, differences of this order are comparable to variations in stellar mass estimates produced by SED fitting assuming different stellar population synthesis models \citep{Conroy13}

It is worth mentioning cosmic variance as a potential source for differences between \sfrmass\ relations from different surveys.
\citet{Whitaker14} investigated this in detail by comparing the \sfrmass\ relation as measured from each of the five CANDELS/3D-HST fields individually (see Appendix B of their paper).
The field-to-field variation they find is comparable to the inter-survey variation that we find.
Given that ZFOURGE covers less area ($\approx 400$ arcmin$^2$ versus $\approx 900$ arcmin$^2$ for 3D-HST) we acknowledge that cosmic variance could explain the differences mentioned in the previous paragraph.
Furthermore, \citet{Tomczak14} quantify cosmic variance among the three ZFOURGE fields in their measurement of the stellar mass function, finding it to range from $8-25$\% over roughly the same mass and \z\ ranges as used in this work.

We do note, however, that for the full sample in Figure \ref{fig:sfr-mass_literature} at $> 10^{10.5}$ \msol\ and $1.5 < z < 2.5$ our SFRs are systematically below the others.
This offset goes away when we recalculate \psiuvir\ from our sample excluding the {\it Herschel} PACS 100 \& 160\um\ photometry but keeping the {\it Spitzer} MIPS 24\um\ photometry.
In Figure \ref{fig:sfr-mass_literature_residuals} we further investigate the impact that the {\it Herschel} PACS imaging has on our stacked SFRs.
For this comparison we only consider star-forming galaxies.
We first perform an internal comparison where we calculate \psiuvir\ both including and excluding the {\it Herschel} stacks.
Interestingly we find a systematic trend wherein the SFRs of higher mass galaxies tend to be overestimated when relying on the MIPS 24\um\ stacks alone for the IR contribution.
This result contrasts what was found for galaxies that are individually detected in the far-IR images which show no strong evidence of a systematic trend (see Fig. \ref{fig:lir-compare}).
This suggests that the fainter, non-detected galaxies have SEDs that are not consistent with their more luminous counterparts \citep{Muzzin10, Wuyts11}.
Nevertheless, this discrepancy is not dominant, but comparable to the level of scatter between surveys noted earlier ($\lesssim 0.2$ dex).

We look into this difference in more detail by making use of the library of IR templates from \citet[hereafter CE01]{Chary01}.
For each mass-\z\ bin we find the individual best-fit CE01 template to the stacked $24-160$\um\ photometry.
Comparing the IR luminosities derived from these best-fit templates to those derived from the W08 template we find a small scatter of $\approx 0.02$ dex with a similar offset in log(\lir/\lsol), and a hint that this offset increases to $\approx 0.1$ dex at $z > 2$. 
However, we don't correct for this effect because these discrepancies are poorly constrained given that the FIR data do not probe the peak of the dust emission.
We also note that this difference is less than the intrinsic uncertainty in individual star-formation rate calibrations (e.g. Bell et al. 2003, 2005).

The differences between the 24\um-derived star-formation rates for 3D-HST and ZFOURGE in Figure \ref{fig:sfr-mass_literature_residuals} are particularly interesting.
These surveys cover the same fields (although ZFOURGE only covers half the area), rely on much of the same public imaging in the optical and IR bands, have had photometry performed using similar methods, and use the same conversions to calculate the star-formation rates from the 2800\AA\ and 24\um\ flux.
Thus the systematic differences in star-formation rates are indicative of the minimal differences that can be expected in inter-survey comparisons.

\subsection{Comparison to Simulations}
\label{sec:simulations}

In Figure \ref{fig:sfr-mass_literature_simulations}, we compare our measured \sfrmass\ relations for the full sample of galaxies with results from the recent Illustris hydrodynamic simulation \citep{Nelson15arxiv} and the Munich semi-analytic galaxy formation model \citep{Henriques14arxiv}.
Gray lines correspond to the mean SFR in bins of stellar mass for each \z\ interval indicated.
Except at \logm\ $> 10.5$, both simulations are consistently in good agreement with each other \citep[for further discussion of this point see][]{Weinmann12}.
In general, the simulations reproduce the roughly constant slope at $M_* \lesssim 10^{10}$ \msol\ albeit with an offset.
This offset ranges between $0.17-0.45$ dex at fixed stellar mass and decreases with \z\ \citep{Sparre15}.
At higher masses, however, Illustris and the Munich galaxy formation model tend to under-predict and over-predict the strength of the turnover at $z < 2$ respectively.

\subsection{Parameterizing the SFR-$M_*$ Relation}
\label{sec:parameterize}

In Figure \ref{fig:sfrmass-parameterization_TOT} we parameterize the \sfrmass\ relation as a function of \z.
For this we adopt the same parameterization as \citet{Lee15}:

\begin{equation}
\label{eq:sfrmass}
\mathrm{log}( \Psi )  =   \mathrm{s}_0  -  \mathrm{log} \left[ 1  +  \left( \frac{M_*}{\mathrm{M}_0} \right)^{-\gamma} \right]
\end{equation} \\[-3mm]

\noindent
where s$_0$ and M$_0$ are in units of log(\msol/yr) and \msol\ respectively.
This function behaves as a power-law of slope $\gamma$ at low masses which asymptotically approaches a peak value s$_0$ above a transitional stellar mass M$_0$.
Originally this parameterization was defined for the \sfrmass\ relation of star-forming galaxies, though we find it works similarly well for the \sfrmass\ relation of all galaxies (star-forming plus \qui) at the \z s and stellar masses considered for this study.
The righthand panels of Figure \ref{fig:sfrmass-parameterization_TOT} show the best-fit parameters vs. \z.
We consider two cases for fitting: ``free $\gamma$'' and ``fixed $\gamma$''.
In the ``free $\gamma$'' case, we allow all three parameters to vary independently for each \z\ bin.
Noticing that $\gamma$ does not show strong evidence for evolution, we perform the ``fixed $\gamma$'' case by refitting with $\gamma$ fixed to its mean value from the ``free $\gamma$'' fits.
We then parameterize the evolution of s$_0$ and M$_0$ with second-order polynomials.

\begin{equation}
\label{eq:parameterization_tot}
\begin{split}
 \mathrm{\bf All \; Galaxie}&\mathrm{\bf s}: \\
 \mathrm{s}_0 \: = & \:\: 0.195  +  1.157 z  -  0.143 z^2\\
 \mathrm{log(M_0)} \: = & \:\:  9.244  +  0.753 z  -  0.090 z^2\\
\gamma \: = & \:\: 1.118
\end{split}
\end{equation}

Evolution in the transition mass is a new discovery, first reported quantitatively at $0.25 < z < 1.3$ by \citet{Lee15} from a study of the COSMOS 2 deg$^2$ field.
Recent results from \citet{Gavazzi15} show that this evolution extends to $z \sim 2.5$ and that extrapolating to $z = 0$ coincides with a break observed in the local \sfrmass\ relation.
Results from the VUDS spectroscopic survey have also found a redshift dependence for the turnover mass \citep{Tasca15}.
However, because we include all galaxies in the analysis presented here the observed evolution of M$_0$ may be a consequence of the increasing population of massive quenched galaxies at low-$z$.
Therefore, we repeat this analysis for a sample of actively star-forming galaxies selected based on \rf\ (U$-$V) and (V$-$J) colors (see Section \ref{sec:sample}). 
Evolution in M$_0$ is still apparent; results are shown in Figure \ref{fig:sfrmass-parameterization_SF}.
In Figure \ref{fig:M0_evolution} we compare our measured values of the turnover mass to those of \citet{Lee15}, \citet{Gavazzi15}, and \citet{Tasca15}.

\begin{equation}
\begin{split}
 \mathrm{\bf Star-Form}&\mathrm{\bf ing}: \\
 \mathrm{s}_0 \: = & \:\: 0.448  +  1.220 z  -  0.174 z^2\\
 \mathrm{log(M_0)} \: = & \:\:  9.458  +  0.865 z  -  0.132 z^2\\
\gamma \: = & \:\: 1.091
\end{split}
\end{equation} \\[-3mm]

We remind the reader that these parameterizations may not extrapolate well outside of the \z\ and/or stellar mass ranges used here.
Nevertheless, our low-mass slope of $\gamma \sim 1$ is consistent with the relative constancy of the low-mass slope $\alpha \sim -1.5$ of the SMF \citep{Tomczak14}; if $\gamma$ deviated significantly from unity then $\alpha$ would be expected to evolve strongly with \z\ \citep{Peng10, Weinmann12, Leja15}.

Figure 8 shows that (s$_0$, M$_0$, $\gamma$) follow similar trends with redshift between the star-forming and total samples indicating that the presence and evolution of the turnover mass is intrinsic to the star-forming population and is not simply due to a growing quiescent population diluting the star-formation rates at the high mass end.
We also point out that the apparent similarity in the fitted parameters does not suggest a redshift-independent quiescent fraction; in fact, the difference between the SFR-M* relations at between the star-forming and the total sample are roughly consistent with the evolution of the quiescent fraction as derived from the Tomczak et al. (2014) mass functions, and assuming negligible star formation in the quiescent galaxies.

\section{Inferring Stellar Mass Growth}

\subsection{Growth of the Stellar Mass Function}
\label{sec:smf-evo}

The growth of galaxies as predicted from the \sfrmass\ relation can be compared directly to the evolution of the stellar mass function.
\citet{Leja15} performed such an analysis using stellar mass functions from \citet{Tomczak14} and \sfrmass\ relations from \citet{Whitaker12} finding that the inferred growth from star-formation greatly over-predicts the observed number densities of galaxies, even on short cosmic timescales ($< 1$ Gyr).
Those authors suggest that the \sfrmass\ relation must have a steeper slope ($\alpha \gtrsim 0.9$) at masses below $10^{10.5}$ \msol\ at $z < 2.5$, which is consistent with measurements from this work and recent literature \citep{Whitaker14, Lee15, Schreiber15}.

Thus, we perform the same comparison using our updated \sfrmass\ relation.
At each stellar mass for a given SMF at a given \z, SFRs are calculated from the parameterized \sfrmass\ relation for all galaxies (Equations \ref{eq:sfrmass} and \ref{eq:parameterization_tot}).
In times steps of 80 Myr each mass bin is shifted by the amount of star formation added to that bin.
SFRs are recalculated at each new time step.
Mass loss due to stellar evolution is accounted for according to Equation 16 of \citet{Moster13}.
Using this technique, we evolve the observed SMF in each redshift bin forward and compare it to the observed SMF in the next redshift bin; results are shown in Figure \ref{fig:smf-evo}

In general, we typically find reasonable agreement at intermediate stellar masses ($10^{10.5} < M_* / M_{\odot} < 10^{11}$). 
At lower masses, however, we find a consistent systematic offset in number density rising to $\approx 0.2-0.3$ dex.
It is important to note that this method {\it does not} incorporate the effect of galaxy-galaxy mergers whereas the observed evolution galaxy SMF necessarily does.
Therefore, the disparity between these two curves inherently includes a signature of merging; in fact, \citet{Drory08} use this difference to constrain galaxy growth rates due to merging. 
Mergers will help to alleviate the discrepancy we observe if these low-mass galaxies are merging with more massive galaxies, thereby reducing their number density.
However this would require between $25-65$\% of these galaxies to merge with a more massive galaxy per Gyr, which substantially exceeds current estimates of galaxy merger rates \citep[e.g.][]{Lotz11, Williams11, Leja15}.
This disagreement thus implies that the SFRs are overestimated and/or the growth of the \citet{Tomczak14} SMF is too slow.
Similar issues were previously discussed by Weinmann et al. (2009) and Leja et al. (2015), but in contrast to the drastic discrepancies between star-formation rates and stellar masses reported by those authors, here we show that using new data sets greatly reduces -- but does not eliminate -- the discrepancies.

\subsection{Empirical Star-Formation Histories}
\label{sec:sfh}

The evolution of the \sfrmass\ relation can be used to infer typical star-formation histories and stellar mass-growth histories for individual galaxies.
However, as seen in the previous subsection, there is tension between the growth of the galaxy population as inferred from the \sfrmass\ relation when comparing to the observed stellar mass function.
Here we explore two different approaches for empirically deriving galaxy mass-growth histories: (1) integrating differential star-formation histories extracted from the \sfrmass\ relation and (2) identifying descendents of high-$z$ galaxies based on number density selected (NDS) samples (Figure \ref{fig:sfhs}).
For the former, we start with the four sets of initial conditions ($z_0$, $M_0$, $\Psi_0$) indicated by the star symbols, where $\Psi_0$ is determined by our SFR parameterizations (Equations \ref{eq:sfrmass} and \ref{eq:parameterization_tot}).
Stellar mass is then incrementally added assuming constant star-formation over small time intervals of $\approx 80$ Myr.
At the end of each time step $\Psi$ adjusted according to the \sfrmass\ parameterization $\Psi(z, M_*)$.
At lower \z s we interpolate between our lowest \z\ \sfrmass\ relation and Equation 13 of \citet{Salim07}.
Mass loss due to stellar evolution is accounted for according to Equation 16 of \citet{Moster13}.

In order to get a rough estimate of the scatter in these SFHs we run 1000 Monte Carlo simulations, resampling $\Psi$ from a log-normal PDF with a mean value given by $\Psi(z, M_*)$ and $\pm$0.3 dex scatter.
The approximate range of mass-growth for each galaxy sample is calculated from the 16\th\ and 84\th\ percentiles of the distribution.

A number of studies have used this technique to estimate galaxy growth histories \citep[e.g.][]{Renzini09, Peng10, Leitner12}.
An important point that should be kept in mind is that in this section we use the \sfrmass\ relation for all galaxies $-$ not just actively star-forming ones $-$ as is appropriate for a comparison to NDS samples.
Some other studies investigate the growth histories of galaxies that remain star-forming without ever quenching \citep{Renzini09, Leitner12}.

Also shown in the middle panel of Figure \ref{fig:sfhs} are mass-growth profiles predicted from the NDS approach.
Using the same initial conditions shown by the star symbols, we calculate the corresponding cumulative co-moving number density from the stellar mass functions of \citet{Tomczak14} as parameterized by \citet{Leja15}.
Note, this parameterization is limited to $z \leq 2.25$, beyond which we interpolate between it and the best-fit Schechter function to the SMF at $2.5 < z < 3$.

Using abundance matching with a dark matter simulation, whereby dark matter halos are assigned stellar masses from observations in a rank ordering fashion, \citet{Behroozi13} have studied the number density evolution of galaxies.
These authors demonstrate that galaxy descendants do not evolve in the same way as their progenitors, mainly due to scatter in dark matter accretion rates of halos.
Thus, \citet{Behroozi13} have provided a numerical recipe for estimating the number density evolution of galaxies, which we use to generate predictions for the number density evolution from the initial conditions given in Figure \ref{fig:sfhs}.
These predictions include the median estimated number density as well as the 68\th\ percentile range.
The hatched regions in Figure \ref{fig:sfhs} show these 68\th\ percentile ranges converted to stellar masses by mapping number densities to the observed stellar mass function.

On average we find that these two approaches agree within their combined 1$\sigma$ confidence intervals.
However there is a common systematic difference wherein the differential SFHs produce a steeper mass-growth rate for the same progenitor galaxy at early times which provides us with a different view of the discrepancy shown in Figure \ref{fig:smf-evo}.
This disparity is illustrated in the rightmost panel of Figure \ref{fig:sfhs} which shows the difference in the mass-growth rates of both techniques from the middle panel.
The differential SFHs build stellar mass more quickly at early times, but then slow down and are eventually overtaken by the NDS growth rates.
We tested a wide range of initial conditions spanning $0.8 \leq z_0 \leq 2.75$ and $8.8 \leq \mathrm{log}( M_0 / M_{\odot} ) \leq 10.5$, always finding this systematic trend.
\citet{Papovich15} find similar results in their analysis of the progenitors of galaxies with present-day masses of the Milky Way and M31 galaxies.
Finally, we have repeated this comparison using the \sfrmass\ parameterizations provided by \citet{Whitaker14} and \citet{Schreiber15} and in both cases we find similar disagreements.
This suggests that the discrepancies between the star-formation rates and mass evolution that were previously reported by Leja et al. (2015) have not been completely resolved using the more up-to-date data sets.

\section{Conclusions}

In this paper we present new measurements of the evolution of the \sfrmass\ relation using deep imaging and high-quality photometric redshifts from the FourStar Galaxy Evolution Survey (\zf) in combination with ancillary far-IR imaging at 24, 100, and 160\um\ from {\it Spitzer} and {\it Herschel}.
Bolometric IR luminosities (\lir), used for calculating obscured star-formation rates, are obtained by scaling the IR spectral template introduced by \citet{Wuyts08} to the $24-160$\um\ photometry.
This luminosity-independent conversion of flux to \lir\ has been shown to be more appropriate than techniques that apply different IR templates for different \lir\ regimes \citep{Muzzin10, Wuyts11}.

Utilizing star-formation rates derived from a UV+IR stacking analysis we examine evolution of the \sfrmass\ relation at $0.5 < z < 4$.
We perform this analysis for all galaxies as well as a sample of actively star-forming galaxies as selected by their \rf\ (U$-$V) and (V$-$J) colors.
In agreement with recent results, we find that SFRs are roughly proportional to stellar mass at low masses ($\lesssim 10^{10.2}$ \msol), but that this trend flattens at higher masses \citep[see also][]{Whitaker14, Lee15, Schreiber15, Tasca15}.
Furthermore, although the evolution of the \sfrmass\ relation is still predominantly in normalization, the slope at high masses ($M_* \gtrsim 10^{10.2}$ \msol) is also changing.
Similar to \citet{Lee15} and \citet{Tasca15} we find that the transition mass at which this flattening occurs evolves with \z; this is true whether or not quenched galaxies are included.
Full parameterizations of the \sfrmass\ relation with respect to \z, $\Psi(z, M_*)$, for both all and star-forming galaxies are presented in Section \ref{sec:parameterize} and shown in Figures \ref{fig:sfrmass-parameterization_TOT} and \ref{fig:sfrmass-parameterization_SF}.

By integrating along the evolving \sfrmass\ sequence we estimate how galaxies should grow due to star-formation.
We find that galaxies with a present-day mass of roughly $10^{10}$ \msol\ have grown in mass by about 10\x\ since $z \sim 1.5$.
In contrast, $10^{11}$ \msol\ galaxies have grown by only about 1.5\x\ since $z \sim 1.5$, but show more rapid evolution at higher redshifts, growing by \twid 15\x\ between over $1.5 < z < 3$.
Furthermore, we find that SFHs rise at early times and fall at late times.
The peak of a galaxy's SFH occurs earlier for galaxies with larger present-day masses; for example galaxies with a present-day stellar mass around $10^{11}$ \msol\ peak at $z \approx 2$, whereas $10^{10}$ \msol\ galaxies peak at $z \approx 0.8$.
Several recent studies have also found evidence in support of rising SFHs in individual galaxies at early times \citep[e.g.][]{Papovich11, Reddy11, Lee11, Abramson15}.

A standard question in galaxy evolution has been whether integrated star-formation rates are consistent with evolution of the global stellar mass density (e.g. Wilkins et al. 2008, Reddy \& Steidel 2009), with recent measurements suggesting that these quantities may be in reasonable -- though not perfect -- agreement (Madau \& Dickinson 2014).
Here we have taken a step further, using new data to ask whether the star-formation rates agree with the mass density evolution in bins of stellar mass.
We use the evolution of the \sfrmass\ relation to predict the growth of galaxies due to star formation which we directly compare to the observed galaxy stellar mass function (Figure \ref{fig:smf-evo}).
At intermediate stellar masses ($10^{10.5} < M_* / M_{\odot} < 10^{11}$) in the redshift range $0.5 < z < 2$ we find reasonable agreement.
However at lower masses the star-formation rates suggest a buildup that is too large in comparison to the evolution of the mass function.
This discrepancy may be partially explained by mergers, in which the lower-mass galaxies merge with more massive ones.
However the merger rates required to resolve this discrepancy are unreasonably large (roughly $25-65$\% per Gyr).
This disagreement, also reported by Leja et al. (2015), implies that the SFRs are overestimated and/or that the growth of the Tomczak et al. (2014) mass function is too slow.

Looking further into the buildup of stellar mass, we use two techniques to extract empirical star-formation and mass-growth histories from observations (see Figure 11).
The first is the method described earlier whereby SFHs are derived by integrating along the evolving \sfrmass\ relation from a set of initial conditions.
The second technique estimates mass-growth histories from measurements of the galaxy stellar mass function using an evolving number density selection (NDS) criterion (Behroozi et al. 2013).
It is worthwhile to note that both techniques provide {\it typical} SFHs along with a rough indication of the scatter, but that individual galaxies may follow very different evolutionary pathways (Kelson 2014; Abramson et al. 2015).
We find that these two methods for deriving galaxy growth histories provide qualitatively similar results, but that they do disagree in detail.
In general we observe a systematic difference wherein the integrated SFHs suggest more rapid mass evolution at higher redshifts than is inferred from the NDS samples.
This disagreement in mass-growth rates reaches to $\gtrsim 0.5$ dex at the highest redshifts that we can probe.
At lower redshifts the NDS predict more rapid evolution; this can be naturally explained by galaxy mergers, and the size of the difference can be taken as a measure of the growth rate due to mergers \citep[e.g.][]{Drory08, Moustakas13}.
Nevertheless, these two approaches on average agree within their combined 1$\sigma$ confidence intervals.

The disagreement at $z \gtrsim 1$ suggests that either our SFRs are overestimated, that the rate of mass-growth inferred from the stellar mass function is underestimated, or both.
Errors in star-formation rate measurements may arise from low-level AGN activity, an incorrect conversion of flux to bolometric UV/IR luminosities, the assumed IMF, and variations in star-forming duty cycles as probed by UV and IR indicators.
Stellar masses were estimated by fitting models to the observed spectral energy distributions (SEDs) of individual galaxies.
Various assumptions that go into the SED-fitting process that are possible sources for systematic errors include smooth exponentially declining SFHs, a single dust screen, a constant IMF, solar metallicity, and assuming that emission lines do not contribute significantly to the observed photometry \citep[for detailed discussions see][]{Marchesini09, Conroy13, Courteau14}.
Analyzing the scatter introduced into the star-formation rate and stellar mass estimates by varying these assumptions would inform the range of possible stellar mass growth histories, but is beyond the scope of this work.

The measurements on which this study is based were performed using high-quality data and standard methods.
Moreover, the use of the same ZFOURGE sample for measuring both the SMF and the \sfrmass\ relations helps provide internal consistency for this work.
Although the broad qualitative agreement that we find in mass-growth histories is encouraging for current studies of galaxy evolution, the disagreements highlight the need to move beyond the simplistic assumptions that underly current data analysis methods.

\acknowledgements

We would like to thank the Mitchell family for their continuing support and in particular the late George P. Mitchell whose commitment to science and astronomy leaves a lasting legacy.
We would also like to thank the Carnegie Observatories and the Las Campanas Observatory for providing the facilities and support necessary to make the ZFOURGE survey possible as well as the Texas A\&M University Brazos HPC cluster which contributed to this project \href{http://brazos.tamu.edu/}{http://brazos.tamu.edu/}.
This work was supported by the National Science Foundation grant AST-1009707.
KG acknowledges support from ARC Grant DP1094370 and ZFOURGE acknowledges The Australian Time Assignment Committee for Magellan time.
GGK was supported by Future Fellowship FT140100933.
Australian access to the Magellan Telescopes was supported through the National Collaborative Research Infrastructure Strategy of the Australian Federal Government.
KEW was supported by an appointment to the NASA Postdoctoral Program at the Goddard Space Flight Center, administered by Oak Ridge Associated Universities through NASA.

\newpage


\begin{longtable*}{c || c c c c | c c c c c}

\caption{\\[0.5mm] \sfrmass\ Relations Data} \\ \hline \hline \\[-2mm]
\label{tab:sfrmass}

 (1) & (2) & (3) & (4) & (5) & (6) & (7) & (8) & (9) & (10) \\[0.7mm]
\z\ & log($M_*$) & log(\luv)$_{\mathrm{all}}$ & log(\lir)$_{\mathrm{all}}$ & log($\Psi$)$_{\mathrm{all}}$ & log(\luv)$_{\mathrm{sf}}$ & log(\lir)$_{\mathrm{sf}}$  & log($\Psi$)$_{\mathrm{sf}}$ & & \\[0.5mm]
range & [\msol] & [\lsol] & [\lsol] & [\msol/yr] & [\lsol] & [\lsol] & [\msol/yr] & $N_{\mathrm{all}}$ & $N_{\mathrm{sf}}$ \\[1mm] \hline \\[-2mm]

$0.50 < z < 0.75$ & 8.625 & $9.05_{-0.02}^{+0.01}$ & $9.18_{-0.11}^{+0.10}$ & $-0.36_{-0.04}^{+0.04}$ & $9.07_{-0.02}^{+0.02}$ & $9.25_{-0.09}^{+0.08}$ & $-0.32_{-0.04}^{+0.03}$ & 493 & 456 \\
    & 8.875 & $9.22_{-0.02}^{+0.04}$ & $9.43_{-0.04}^{+0.05}$ & $-0.16_{-0.02}^{+0.03}$ & $9.28_{-0.01}^{+0.02}$ & $9.46_{-0.05}^{+0.04}$ & $-0.11_{-0.02}^{+0.02}$ & 391 & 343 \\
    & 9.125 & $9.38_{-0.02}^{+0.06}$ & $9.75_{-0.04}^{+0.04}$ & $0.08_{-0.03}^{+0.03}$ & $9.47_{-0.02}^{+0.02}$ & $9.82_{-0.03}^{+0.04}$ & $0.15_{-0.02}^{+0.03}$ & 300 & 260 \\
    & 9.375 & $9.50_{-0.02}^{+0.06}$ & $10.04_{-0.04}^{+0.03}$ & $0.29_{-0.03}^{+0.02}$ & $9.57_{-0.03}^{+0.03}$ & $10.09_{-0.03}^{+0.03}$ & $0.35_{-0.02}^{+0.02}$ & 261 & 234 \\
    & 9.625 & $9.49_{-0.04}^{+0.10}$ & $10.42_{-0.02}^{+0.02}$ & $0.55_{-0.02}^{+0.02}$ & $9.64_{-0.06}^{+0.04}$ & $10.47_{-0.02}^{+0.02}$ & $0.63_{-0.02}^{+0.02}$ & 203 & 175 \\
    & 9.875 & $9.36_{-0.08}^{+0.03}$ & $10.56_{-0.02}^{+0.02}$ & $0.66_{-0.02}^{+0.02}$ & $9.51_{-0.05}^{+0.03}$ & $10.70_{-0.02}^{+0.02}$ & $0.79_{-0.02}^{+0.02}$ & 146 & 111 \\
    & 10.125 & $9.31_{-0.07}^{+0.07}$ & $10.64_{-0.03}^{+0.04}$ & $0.72_{-0.03}^{+0.03}$ & $9.60_{-0.05}^{+0.06}$ & $10.83_{-0.03}^{+0.02}$ & $0.92_{-0.02}^{+0.02}$ & 147 & 93 \\
    & 10.375 & $9.39_{-0.04}^{+0.03}$ & $10.74_{-0.06}^{+0.04}$ & $0.81_{-0.05}^{+0.04}$ & $9.65_{-0.05}^{+0.07}$ & $11.01_{-0.05}^{+0.05}$ & $1.08_{-0.05}^{+0.04}$ & 101 & 56 \\
    & 10.625 & $9.49_{-0.02}^{+0.04}$ & $10.66_{-0.08}^{+0.06}$ & $0.75_{-0.07}^{+0.06}$ & $9.67_{-0.08}^{+0.05}$ & $11.01_{-0.07}^{+0.07}$ & $1.09_{-0.08}^{+0.06}$ & 67 & 30 \\
    & 10.875 & $9.66_{-0.03}^{+0.02}$ & $10.57_{-0.07}^{+0.06}$ & $0.71_{-0.05}^{+0.05}$ & $9.74_{-0.07}^{+0.04}$ & $10.79_{-0.12}^{+0.08}$ & $0.91_{-0.10}^{+0.07}$ & 46 & 27 \\
    & 11.125 & $9.73_{-0.04}^{+0.05}$ & $10.46_{-0.16}^{+0.10}$ & $0.65_{-0.11}^{+0.09}$ & $9.84_{-0.06}^{+0.05}$ & $10.81_{-0.09}^{+0.06}$ & $0.94_{-0.06}^{+0.07}$ & 18 & 8 \\[1mm] \hline \\[-2mm]

$0.75 < z < 1.00$ & 8.625 & $9.25_{-0.02}^{+0.01}$ & $8.57_{-0.38}^{+0.23}$ & $-0.33_{-0.04}^{+0.04}$ & $9.27_{-0.02}^{+0.00}$ & $8.61_{-0.26}^{+0.24}$ & $-0.31_{-0.03}^{+0.03}$ & 599 & 583 \\
    & 8.875 & $9.42_{-0.02}^{+0.01}$ & $9.48_{-0.10}^{+0.08}$ & $-0.02_{-0.03}^{+0.03}$ & $9.43_{-0.01}^{+0.02}$ & $9.52_{-0.10}^{+0.07}$ & $0.00_{-0.03}^{+0.03}$ & 477 & 452 \\
    & 9.125 & $9.59_{-0.03}^{+0.02}$ & $9.93_{-0.06}^{+0.03}$ & $0.27_{-0.03}^{+0.02}$ & $9.62_{-0.02}^{+0.03}$ & $9.95_{-0.05}^{+0.05}$ & $0.30_{-0.03}^{+0.02}$ & 376 & 351 \\
    & 9.375 & $9.63_{-0.04}^{+0.02}$ & $10.27_{-0.02}^{+0.03}$ & $0.48_{-0.02}^{+0.02}$ & $9.66_{-0.02}^{+0.04}$ & $10.31_{-0.03}^{+0.02}$ & $0.52_{-0.02}^{+0.02}$ & 298 & 268 \\
    & 9.625 & $9.79_{-0.05}^{+0.02}$ & $10.48_{-0.03}^{+0.02}$ & $0.68_{-0.02}^{+0.02}$ & $9.83_{-0.02}^{+0.04}$ & $10.54_{-0.03}^{+0.02}$ & $0.73_{-0.02}^{+0.02}$ & 202 & 178 \\
    & 9.875 & $9.66_{-0.08}^{+0.03}$ & $10.75_{-0.02}^{+0.02}$ & $0.86_{-0.02}^{+0.02}$ & $9.78_{-0.06}^{+0.07}$ & $10.82_{-0.03}^{+0.02}$ & $0.94_{-0.02}^{+0.02}$ & 162 & 137 \\
    & 10.125 & $9.53_{-0.07}^{+0.04}$ & $10.91_{-0.03}^{+0.03}$ & $0.99_{-0.03}^{+0.03}$ & $9.74_{-0.05}^{+0.15}$ & $11.07_{-0.02}^{+0.02}$ & $1.15_{-0.02}^{+0.02}$ & 128 & 85 \\
    & 10.375 & $9.52_{-0.07}^{+0.04}$ & $10.89_{-0.05}^{+0.03}$ & $0.96_{-0.04}^{+0.04}$ & $9.74_{-0.04}^{+0.04}$ & $11.09_{-0.04}^{+0.03}$ & $1.17_{-0.04}^{+0.03}$ & 107 & 64 \\
    & 10.625 & $9.72_{-0.06}^{+0.03}$ & $11.02_{-0.05}^{+0.04}$ & $1.11_{-0.04}^{+0.03}$ & $9.83_{-0.07}^{+0.10}$ & $11.28_{-0.02}^{+0.02}$ & $1.35_{-0.02}^{+0.03}$ & 77 & 44 \\
    & 10.875 & $9.66_{-0.02}^{+0.05}$ & $10.94_{-0.06}^{+0.06}$ & $1.03_{-0.06}^{+0.05}$ & $9.93_{-0.15}^{+0.07}$ & $11.20_{-0.09}^{+0.09}$ & $1.29_{-0.09}^{+0.08}$ & 42 & 22 \\
    & 11.125 & $9.89_{-0.09}^{+0.02}$ & $10.92_{-0.08}^{+0.07}$ & $1.04_{-0.06}^{+0.06}$ & $9.97_{-0.06}^{+0.06}$ & $11.19_{-0.14}^{+0.08}$ & $1.28_{-0.11}^{+0.08}$ & 23 & 12 \\[1mm] \hline \\[-2mm]

$1.00 < z < 1.25$ & 8.625 & $9.40_{-0.02}^{+0.01}$ & $8.98_{-0.62}^{+0.29}$ & $-0.15_{-0.05}^{+0.04}$ & $9.40_{-0.01}^{+0.01}$ & $9.00_{-0.46}^{+0.17}$ & $-0.15_{-0.05}^{+0.04}$ & 371 & 368 \\
    & 8.875 & $9.55_{-0.02}^{+0.01}$ & $9.33_{-0.20}^{+0.11}$ & $0.04_{-0.04}^{+0.03}$ & $9.55_{-0.02}^{+0.01}$ & $9.32_{-0.12}^{+0.16}$ & $0.04_{-0.04}^{+0.04}$ & 379 & 376 \\
    & 9.125 & $9.76_{-0.02}^{+0.01}$ & $10.04_{-0.05}^{+0.08}$ & $0.41_{-0.04}^{+0.04}$ & $9.77_{-0.01}^{+0.01}$ & $10.05_{-0.08}^{+0.07}$ & $0.42_{-0.05}^{+0.03}$ & 284 & 282 \\
    & 9.375 & $9.84_{-0.03}^{+0.01}$ & $10.36_{-0.03}^{+0.05}$ & $0.62_{-0.03}^{+0.02}$ & $9.85_{-0.02}^{+0.02}$ & $10.37_{-0.03}^{+0.04}$ & $0.63_{-0.03}^{+0.03}$ & 248 & 239 \\
    & 9.625 & $9.93_{-0.05}^{+0.05}$ & $10.57_{-0.04}^{+0.03}$ & $0.78_{-0.03}^{+0.03}$ & $9.99_{-0.06}^{+0.03}$ & $10.61_{-0.04}^{+0.04}$ & $0.83_{-0.03}^{+0.03}$ & 157 & 146 \\
    & 9.875 & $9.78_{-0.03}^{+0.05}$ & $10.83_{-0.03}^{+0.02}$ & $0.95_{-0.02}^{+0.02}$ & $9.85_{-0.06}^{+0.06}$ & $10.88_{-0.02}^{+0.02}$ & $1.00_{-0.02}^{+0.02}$ & 113 & 99 \\
    & 10.125 & $9.72_{-0.05}^{+0.07}$ & $11.03_{-0.02}^{+0.03}$ & $1.11_{-0.02}^{+0.02}$ & $9.85_{-0.05}^{+0.05}$ & $11.14_{-0.02}^{+0.02}$ & $1.22_{-0.02}^{+0.02}$ & 123 & 95 \\
    & 10.375 & $9.60_{-0.05}^{+0.05}$ & $11.10_{-0.03}^{+0.03}$ & $1.17_{-0.03}^{+0.03}$ & $9.85_{-0.07}^{+0.04}$ & $11.31_{-0.04}^{+0.03}$ & $1.38_{-0.04}^{+0.03}$ & 92 & 57 \\
    & 10.625 & $9.74_{-0.02}^{+0.02}$ & $11.26_{-0.03}^{+0.03}$ & $1.32_{-0.03}^{+0.03}$ & $9.82_{-0.08}^{+0.21}$ & $11.39_{-0.04}^{+0.03}$ & $1.46_{-0.03}^{+0.03}$ & 83 & 59 \\
    & 10.875 & $9.86_{-0.04}^{+0.08}$ & $11.28_{-0.06}^{+0.06}$ & $1.35_{-0.06}^{+0.06}$ & $9.96_{-0.01}^{+0.07}$ & $11.42_{-0.08}^{+0.08}$ & $1.49_{-0.08}^{+0.07}$ & 36 & 22 \\
    & 11.125 & $9.89_{-0.01}^{+0.00}$ & $11.61_{-0.09}^{+0.05}$ & $1.55_{-0.07}^{+0.06}$ & $9.93_{-0.06}^{+0.09}$ & $11.76_{-0.05}^{+0.06}$ & $1.71_{-0.05}^{+0.06}$ & 12 & 8 \\[1mm] \hline \\[-2mm]

$1.25 < z < 1.50$ & 8.625 & $9.52_{-0.01}^{+0.02}$ & $8.18_{-0.93}^{+0.95}$ & $-0.09_{-0.06}^{+0.05}$ & $9.52_{-0.01}^{+0.02}$ & $8.25_{-0.63}^{+0.71}$ & $-0.09_{-0.07}^{+0.06}$ & 287 & 286 \\
    & 8.875 & $9.65_{-0.01}^{+0.01}$ & $9.41_{-0.17}^{+0.13}$ & $0.13_{-0.04}^{+0.03}$ & $9.65_{-0.01}^{+0.01}$ & $9.41_{-0.19}^{+0.13}$ & $0.13_{-0.04}^{+0.03}$ & 429 & 426 \\
    & 9.125 & $9.80_{-0.02}^{+0.01}$ & $10.07_{-0.13}^{+0.07}$ & $0.45_{-0.05}^{+0.04}$ & $9.81_{-0.01}^{+0.01}$ & $10.09_{-0.11}^{+0.09}$ & $0.46_{-0.05}^{+0.05}$ & 354 & 348 \\
    & 9.375 & $9.94_{-0.03}^{+0.03}$ & $10.37_{-0.06}^{+0.05}$ & $0.67_{-0.03}^{+0.03}$ & $9.94_{-0.03}^{+0.03}$ & $10.37_{-0.06}^{+0.05}$ & $0.67_{-0.03}^{+0.03}$ & 269 & 265 \\
    & 9.625 & $9.98_{-0.02}^{+0.02}$ & $10.74_{-0.03}^{+0.03}$ & $0.92_{-0.02}^{+0.02}$ & $9.99_{-0.02}^{+0.04}$ & $10.76_{-0.04}^{+0.02}$ & $0.93_{-0.03}^{+0.02}$ & 205 & 197 \\
    & 9.875 & $10.01_{-0.04}^{+0.04}$ & $10.88_{-0.04}^{+0.03}$ & $1.03_{-0.03}^{+0.02}$ & $10.04_{-0.02}^{+0.04}$ & $10.93_{-0.03}^{+0.02}$ & $1.07_{-0.03}^{+0.02}$ & 151 & 141 \\
    & 10.125 & $9.90_{-0.05}^{+0.05}$ & $11.17_{-0.03}^{+0.03}$ & $1.26_{-0.03}^{+0.03}$ & $10.0_{-0.06}^{+0.04}$ & $11.23_{-0.03}^{+0.03}$ & $1.32_{-0.03}^{+0.03}$ & 148 & 128 \\
    & 10.375 & $9.70_{-0.02}^{+0.07}$ & $11.27_{-0.03}^{+0.04}$ & $1.33_{-0.04}^{+0.04}$ & $9.83_{-0.05}^{+0.06}$ & $11.40_{-0.04}^{+0.03}$ & $1.46_{-0.04}^{+0.03}$ & 121 & 90 \\
    & 10.625 & $9.82_{-0.06}^{+0.06}$ & $11.30_{-0.06}^{+0.06}$ & $1.37_{-0.06}^{+0.05}$ & $9.96_{-0.06}^{+0.02}$ & $11.56_{-0.05}^{+0.04}$ & $1.63_{-0.05}^{+0.04}$ & 77 & 45 \\
    & 10.875 & $9.98_{-0.05}^{+0.03}$ & $11.54_{-0.11}^{+0.09}$ & $1.61_{-0.11}^{+0.09}$ & $10.03_{-0.03}^{+0.05}$ & $11.78_{-0.06}^{+0.07}$ & $1.83_{-0.06}^{+0.07}$ & 56 & 37 \\
    & 11.125 & $10.0_{-0.01}^{+0.04}$ & $11.59_{-0.09}^{+0.07}$ & $1.65_{-0.09}^{+0.07}$ & $9.97_{-0.04}^{+0.24}$ & $11.88_{-0.05}^{+0.05}$ & $1.93_{-0.05}^{+0.05}$ & 14 & 8 \\[1mm] \hline \\[-2mm]

$1.50 < z < 2.00$ & 8.875 & $9.76_{-0.01}^{+0.01}$ & $8.51_{-0.55}^{+0.71}$ & $0.11_{-0.06}^{+0.06}$ & $9.76_{-0.01}^{+0.01}$ & $8.53_{-0.54}^{+0.68}$ & $0.11_{-0.08}^{+0.06}$ & 645 & 644 \\
    & 9.125 & $9.86_{-0.01}^{+0.01}$ & $10.22_{-0.09}^{+0.10}$ & $0.55_{-0.05}^{+0.05}$ & $9.86_{-0.01}^{+0.02}$ & $10.23_{-0.09}^{+0.09}$ & $0.56_{-0.06}^{+0.05}$ & 752 & 747 \\
    & 9.375 & $10.03_{-0.01}^{+0.01}$ & $10.47_{-0.05}^{+0.04}$ & $0.76_{-0.03}^{+0.03}$ & $10.03_{-0.01}^{+0.02}$ & $10.47_{-0.05}^{+0.04}$ & $0.76_{-0.03}^{+0.03}$ & 616 & 607 \\
    & 9.625 & $10.10_{-0.03}^{+0.01}$ & $10.87_{-0.02}^{+0.03}$ & $1.04_{-0.02}^{+0.02}$ & $10.11_{-0.01}^{+0.02}$ & $10.90_{-0.02}^{+0.03}$ & $1.07_{-0.02}^{+0.02}$ & 438 & 416 \\
    & 9.875 & $10.06_{-0.02}^{+0.02}$ & $11.04_{-0.03}^{+0.03}$ & $1.17_{-0.03}^{+0.03}$ & $10.12_{-0.03}^{+0.04}$ & $11.10_{-0.02}^{+0.02}$ & $1.23_{-0.02}^{+0.02}$ & 316 & 287 \\
    & 10.125 & $9.97_{-0.02}^{+0.05}$ & $11.30_{-0.02}^{+0.02}$ & $1.38_{-0.02}^{+0.02}$ & $10.10_{-0.05}^{+0.04}$ & $11.38_{-0.02}^{+0.02}$ & $1.47_{-0.02}^{+0.02}$ & 239 & 199 \\
    & 10.375 & $9.74_{-0.05}^{+0.04}$ & $11.43_{-0.04}^{+0.02}$ & $1.48_{-0.03}^{+0.03}$ & $9.88_{-0.03}^{+0.07}$ & $11.56_{-0.02}^{+0.03}$ & $1.62_{-0.02}^{+0.02}$ & 179 & 133 \\
    & 10.625 & $9.84_{-0.04}^{+0.02}$ & $11.51_{-0.03}^{+0.03}$ & $1.57_{-0.03}^{+0.03}$ & $10.00_{-0.09}^{+0.02}$ & $11.75_{-0.03}^{+0.02}$ & $1.80_{-0.03}^{+0.03}$ & 142 & 94 \\
    & 10.875 & $10.06_{-0.06}^{+0.03}$ & $11.61_{-0.05}^{+0.04}$ & $1.67_{-0.05}^{+0.04}$ & $10.07_{-0.06}^{+0.02}$ & $11.80_{-0.05}^{+0.04}$ & $1.86_{-0.04}^{+0.04}$ & 119 & 83 \\
    & 11.125 & $10.15_{-0.06}^{+0.05}$ & $11.63_{-0.08}^{+0.07}$ & $1.70_{-0.07}^{+0.07}$ & $10.37_{-0.23}^{+0.02}$ & $11.91_{-0.06}^{+0.06}$ & $1.98_{-0.06}^{+0.06}$ & 30 & 19 \\[1mm] \hline \\[-2mm]

$2.00 < z < 2.50$ & 9.125 & $10.04_{-0.01}^{+0.01}$ & $9.76_{-0.46}^{+0.23}$ & $0.51_{-0.06}^{+0.06}$ & $10.04_{-0.01}^{+0.01}$ & $9.76_{-0.50}^{+0.29}$ & $0.51_{-0.06}^{+0.07}$ & 475 & 475 \\
    & 9.375 & $10.14_{-0.01}^{+0.01}$ & $10.42_{-0.11}^{+0.09}$ & $0.79_{-0.05}^{+0.05}$ & $10.14_{-0.01}^{+0.01}$ & $10.41_{-0.15}^{+0.11}$ & $0.79_{-0.07}^{+0.06}$ & 514 & 513 \\
    & 9.625 & $10.27_{-0.01}^{+0.02}$ & $10.79_{-0.07}^{+0.05}$ & $1.05_{-0.04}^{+0.03}$ & $10.27_{-0.01}^{+0.02}$ & $10.79_{-0.05}^{+0.04}$ & $1.05_{-0.03}^{+0.03}$ & 403 & 400 \\
    & 9.875 & $10.36_{-0.02}^{+0.02}$ & $11.19_{-0.04}^{+0.03}$ & $1.35_{-0.03}^{+0.03}$ & $10.38_{-0.02}^{+0.04}$ & $11.20_{-0.04}^{+0.04}$ & $1.36_{-0.03}^{+0.03}$ & 250 & 236 \\
    & 10.125 & $10.28_{-0.06}^{+0.08}$ & $11.28_{-0.03}^{+0.03}$ & $1.40_{-0.04}^{+0.03}$ & $10.37_{-0.02}^{+0.05}$ & $11.35_{-0.03}^{+0.03}$ & $1.47_{-0.03}^{+0.02}$ & 197 & 173 \\
    & 10.375 & $10.17_{-0.07}^{+0.06}$ & $11.54_{-0.05}^{+0.05}$ & $1.61_{-0.05}^{+0.05}$ & $10.25_{-0.04}^{+0.05}$ & $11.64_{-0.05}^{+0.04}$ & $1.71_{-0.04}^{+0.04}$ & 124 & 103 \\
    & 10.625 & $9.95_{-0.06}^{+0.04}$ & $11.58_{-0.06}^{+0.04}$ & $1.64_{-0.05}^{+0.04}$ & $9.95_{-0.07}^{+0.09}$ & $11.72_{-0.04}^{+0.04}$ & $1.78_{-0.04}^{+0.04}$ & 107 & 81 \\
    & 10.875 & $10.12_{-0.05}^{+0.05}$ & $11.78_{-0.03}^{+0.04}$ & $1.83_{-0.03}^{+0.03}$ & $10.13_{-0.06}^{+0.05}$ & $11.94_{-0.03}^{+0.03}$ & $1.99_{-0.03}^{+0.03}$ & 79 & 55 \\
    & 11.125 & $10.16_{-0.06}^{+0.03}$ & $11.93_{-0.04}^{+0.07}$ & $1.98_{-0.05}^{+0.06}$ & $10.12_{-0.04}^{+0.06}$ & $12.04_{-0.04}^{+0.04}$ & $2.09_{-0.04}^{+0.04}$ & 36 & 29 \\[1mm] \hline \\[-2mm]

$2.50 < z < 3.00$ & 9.125 & $10.23_{-0.02}^{+0.01}$ & $9.35_{-0.21}^{+0.41}$ & $0.57_{-0.06}^{+0.07}$ & $10.23_{-0.01}^{+0.02}$ & $9.35_{-0.73}^{+0.55}$ & $0.58_{-0.06}^{+0.06}$ & 242 & 240 \\
    & 9.375 & $10.29_{-0.01}^{+0.01}$ & $9.35_{-0.59}^{+0.41}$ & $0.69_{-0.03}^{+0.04}$ & $10.29_{-0.01}^{+0.01}$ & $9.36_{-0.56}^{+0.41}$ & $0.69_{-0.06}^{+0.05}$ & 327 & 323 \\
    & 9.625 & $10.41_{-0.02}^{+0.02}$ & $10.99_{-0.12}^{+0.09}$ & $1.22_{-0.08}^{+0.05}$ & $10.41_{-0.02}^{+0.02}$ & $10.98_{-0.14}^{+0.07}$ & $1.22_{-0.08}^{+0.06}$ & 269 & 265 \\
    & 9.875 & $10.52_{-0.06}^{+0.02}$ & $11.27_{-0.10}^{+0.08}$ & $1.45_{-0.07}^{+0.06}$ & $10.52_{-0.03}^{+0.04}$ & $11.27_{-0.07}^{+0.07}$ & $1.45_{-0.05}^{+0.06}$ & 151 & 150 \\
    & 10.125 & $10.39_{-0.03}^{+0.04}$ & $11.67_{-0.06}^{+0.05}$ & $1.76_{-0.06}^{+0.05}$ & $10.42_{-0.04}^{+0.04}$ & $11.68_{-0.05}^{+0.04}$ & $1.76_{-0.04}^{+0.04}$ & 104 & 97 \\
    & 10.375 & $10.14_{-0.06}^{+0.03}$ & $11.66_{-0.12}^{+0.07}$ & $1.73_{-0.11}^{+0.07}$ & $10.29_{-0.11}^{+0.11}$ & $11.79_{-0.10}^{+0.05}$ & $1.86_{-0.08}^{+0.06}$ & 86 & 67 \\
    & 10.625 & $9.92_{-0.04}^{+0.16}$ & $11.86_{-0.05}^{+0.05}$ & $1.91_{-0.05}^{+0.05}$ & $10.03_{-0.15}^{+0.05}$ & $11.95_{-0.04}^{+0.04}$ & $2.00_{-0.04}^{+0.04}$ & 66 & 56 \\
    & 10.875 & $9.95_{-0.05}^{+0.18}$ & $11.98_{-0.06}^{+0.06}$ & $2.03_{-0.06}^{+0.06}$ & $9.89_{-0.03}^{+0.06}$ & $12.09_{-0.06}^{+0.05}$ & $2.13_{-0.05}^{+0.05}$ & 29 & 24 \\
    & 11.125 & $10.23_{-0.08}^{+0.22}$ & $12.32_{-0.10}^{+0.08}$ & $2.37_{-0.10}^{+0.08}$ & $10.32_{-0.14}^{+0.13}$ & $12.35_{-0.09}^{+0.08}$ & $2.40_{-0.09}^{+0.07}$ & 15 & 14 \\[1mm] \hline \\[-2mm]

$3.00 < z < 4.00$ & 9.625 & $10.42_{-0.01}^{+0.01}$ & $10.64_{-0.25}^{+0.17}$ & $1.04_{-0.10}^{+0.08}$ & $10.43_{-0.01}^{+0.01}$ & $10.64_{-0.26}^{+0.11}$ & $1.05_{-0.09}^{+0.07}$ & 256 & 253 \\
    & 9.875 & $10.53_{-0.01}^{+0.04}$ & $11.48_{-0.09}^{+0.08}$ & $1.61_{-0.07}^{+0.06}$ & $10.54_{-0.03}^{+0.03}$ & $11.47_{-0.09}^{+0.06}$ & $1.61_{-0.07}^{+0.06}$ & 166 & 161 \\
    & 10.125 & $10.55_{-0.05}^{+0.03}$ & $11.63_{-0.06}^{+0.07}$ & $1.74_{-0.05}^{+0.06}$ & $10.58_{-0.02}^{+0.04}$ & $11.67_{-0.08}^{+0.06}$ & $1.78_{-0.06}^{+0.06}$ & 122 & 111 \\
    & 10.375 & $10.41_{-0.07}^{+0.08}$ & $11.83_{-0.07}^{+0.06}$ & $1.90_{-0.06}^{+0.06}$ & $10.49_{-0.10}^{+0.08}$ & $11.86_{-0.06}^{+0.05}$ & $1.94_{-0.06}^{+0.05}$ & 65 & 58 \\
    & 10.625 & $10.34_{-0.16}^{+0.21}$ & $12.06_{-0.12}^{+0.12}$ & $2.11_{-0.12}^{+0.11}$ & $10.54_{-0.30}^{+0.10}$ & $12.14_{-0.11}^{+0.11}$ & $2.20_{-0.11}^{+0.09}$ & 34 & 28 \\
    & 10.875 & $10.14_{-0.10}^{+0.17}$ & $12.20_{-0.07}^{+0.07}$ & $2.24_{-0.07}^{+0.07}$ & $10.07_{-0.06}^{+0.18}$ & $12.26_{-0.08}^{+0.06}$ & $2.31_{-0.07}^{+0.06}$ & 30 & 24 \\
    & 11.125 & $10.22_{-0.13}^{+0.21}$ & $12.22_{-0.13}^{+0.07}$ & $2.27_{-0.12}^{+0.07}$ & $10.22_{-0.12}^{+0.34}$ & $12.33_{-0.09}^{+0.09}$ & $2.37_{-0.09}^{+0.08}$ & 14 & 12 \\
    & 11.375 & $10.16_{-0.07}^{+0.03}$ & $12.43_{-0.05}^{+0.04}$ & $2.47_{-0.05}^{+0.04}$ & $10.16_{-0.07}^{+0.03}$ & $12.48_{-0.07}^{+0.05}$ & $2.52_{-0.07}^{+0.05}$ & 7 & 7 \\[1mm] \hline

\multicolumn{10}{l}{{\bf Notes:} A downloadable ascii version of this table will be hosted at http://zfourge.tamu.edu/}

\end{longtable*}

\nocite{*}
\bibliographystyle{apj}
\bibliography{bibliography}

\end{document}